\documentclass[11pt]{article}
\usepackage{graphicx,color}  
\usepackage{float}
\usepackage{dcolumn}   
\usepackage{bm}        
\usepackage{amssymb} 
\usepackage{amsmath}
\usepackage{braket}
\usepackage{cancel}
\usepackage{subfigure}
\usepackage{fullpage}

\newcommand{\ka}{\kappa}
\newcommand{\del}{\delta}

\newcommand{\pa}{\partial}

\def\a{\alpha}
\def\b{\beta}

\def\m{\mu}
\def\n{\nu}

\newcommand{\cS}{\mathcal{S}}
 \newcommand{\beq}{\begin{equation}}
\newcommand{\eeq}{\end{equation}}
\newcommand{\beqa}{\begin{eqnarray}}
\newcommand{\eeqa}{\end{eqnarray}}
\newcommand{\be}{\begin{equation}}
\newcommand{\ee}{\end{equation}}
\newcommand{\bea}{\begin{eqnarray}}
\newcommand{\eea}{\end{eqnarray}}

\newcommand{\nc}{\newcommand}
\nc{\rnc}{\renewcommand}
\nc{\D}{\partial}
\nc{\K}{\kappa}
\nc{\bK}{\bar{\K}}
\nc{\bN}{\bar{N}}
\nc{\bq}{\bar{q}}
\nc{\vbq}{\vec{\bar{q}}}
\nc{\g}{\gamma}
\nc{\lrarrow}{\leftrightarrow}
\nc{\rg}{\sqrt{g}}
\nc{\de}{\delta}
\let\d=\delta
\nc{\nn}{\nonumber}
\rnc{\(}{\left(}
\rnc{\)}{\right)}
\nc{\q}{\vec{q}}
\nc{\x}{\vec{x}}
\nc{\ep}{\epsilon}
\nc{\tto}{\rightarrow}
\rnc{\inf}{\infty}
\rnc{\Re}{\mathrm{Re}}
\rnc{\Im}{\mathrm{Im}}
\nc{\z}{\zeta}
\nc{\mA}{\mathcal{A}}
\nc{\mB}{\mathcal{B}}
\nc{\mC}{\mathcal{C}}
\nc{\mD}{\mathcal{D}}
\nc{\mN}{\mathcal{N}}
\rnc{\H}{\mathcal{H}}
\rnc{\L}{\mathcal{L}}
\nc{\<}{\langle}
\rnc{\>}{\rangle}
\nc{\fnl}{f_{NL}}
\nc{\gnl}{g_{NL}}
\nc{\fnleq}{f_{NL}^{equil.}}
\nc{\fnlloc}{f_{NL}^{local}}
\nc{\vphi}{\varphi}
\nc{\Lie}{\pounds}
\nc{\half}{\frac{1}{2}}
\nc{\bOmega}{\bar{\Omega}}
\nc{\bLambda}{\bar{\Lambda}}
\nc{\dN}{\delta N}
\nc{\gYM}{g_{\mathrm{YM}}}
\nc{\geff}{g_{\mathrm{eff}}}
\nc{\tr}{\mathrm{tr}}
\nc{\oa}{\stackrel{\leftrightarrow}}
\nc{\IR}{{\rm IR}}
\nc{\UV}{{\rm UV}}
\nc{\la}{\lambda}
\nc{\veps}{\varepsilon}
\begin{document}
\titlepage
	\begin{center}
		\vspace{1.5cm}
		{\Large \bf   Parity-Violating CFT and the Gravitational Chiral Anomaly  \\}
		\vspace{0.3cm}
		
		\vspace{0.3cm}
		
		\vspace{1cm}
		{\bf Claudio Corian\`o$^{(1)}$, Stefano Lionetti\bf $^{(1)}$ and Matteo Maria Maglio$^{(2)}$ \\}
		\vspace{1cm}
		{\it  $^{(1)}$Dipartimento di Matematica e Fisica, Universit\`{a} del Salento \\
			and INFN Sezione di Lecce, Via Arnesano 73100 Lecce, Italy\\
			and National Center for HPC, Big Data and Quantum Computing\\}
		\vspace{0.5cm}
		{\it  $^{(2)}$ Institute for Theoretical Physics (ITP), University of Heidelberg\\
			Philosophenweg 16, 69120 Heidelberg, Germany}

		\begin{abstract}
	We illustrate how the conformal Ward identities (CWI) and the gravitational chiral anomaly completely determine the structure of the $\langle TTJ_{5}\rangle$ (graviton-graviton-chiral gauge current) correlator in momentum space. This analysis extends our previous results on the anomaly vertices $\langle AVV\rangle$ and $\langle AAA\rangle$, as well as the $\langle TJJ\rangle$ parity-odd conformal anomaly vertex in general CFTs. The $\langle TTJ_{5}\rangle$ plays a fundamental role in the analysis of the conformal backreaction in early universe cosmology, affecting the particle content and the evolution of the primordial plasma. Our approach is nonperturbative and not Lagrangian-based, requiring the inclusion of a single anomaly pole in the solution of the anomaly constraint. The pole and its residue, along with the CWIs, determine the entire correlator in all of its sectors (longitudinal/transverse), all of which are proportional to the same anomaly coefficient. The method does not rely on a specific expression of the CP-odd anomalous current, which in free field theory can be represented either by a bilinear fermion current or by a gauge-dependent Chern-Simons current; it relies solely on the symmetry constraints. We compute the correlator perturbatively at one-loop in free field theory and verify its exact agreement with the non-perturbative result. A comparison with the perturbative analysis confirms the presence of a sum rule satisfied by the correlator, similar to the parity-even $\langle TJJ\rangle$ and the chiral $\langle AVV\rangle$.	

 		\end{abstract}
	\end{center}
\newpage
\setcounter{page}{1}
\section{Introduction}
 The original approach to identifying correlation functions in conformal field theories (CFTs) has traditionally been formulated using coordinate space methods, both for scalar and tensor correlators.
In the presence of anomalies, the solutions of the corresponding conformal Ward identities (CWIs) have been obtained by partitioning the domain of definition of each correlator into nonlocal and contact contributions. The equations are initially solved for the regions in which the external coordinate points of the correlators are all noncoincident. Anomalous corrections, which arise when all the points coincide, are manually added by including additional local terms with support defined by products of delta functions.\\ This methodology was pioneered in groundbreaking works \cite{Osborn:1993cr, Erdmenger:1996yc}, and it was successfully applied to correlators containing the stress-energy tensor $(T)$ and conserved vector currents $(J)$, specifically the $\langle TTT\rangle$ and the $\langle TJJ\rangle $ correlators, respectively.
On the other hand, investigations of the conformal constraints in momentum space, in the presence of conformal anomalies, are more recent. This approach has been explored in several works \cite{Coriano:2013jba, Bzowski:2013sza, Bzowski:2015pba,Bzowski:2017poo, Bzowski:2018fql,Marotta:2022jrp}, using a general formulation, and it has been further examined in perturbation theory \cite{Coriano:2018bbe, Coriano:2018zdo, Coriano:2018bsy} for correlation functions such as the $\langle TJJ\rangle$ and the $\langle TTT\rangle$, using free field theory realizations. The analysis of 4-point functions in both generic CFTs and in free field realizations has been discussed in \cite{Coriano:2019nkw, Coriano:2022jkn}.
These analyses have predominantly focused on the parity-even sector, including the contribution of the conformal anomaly. In contrast, investigations into anomaly-free correlators of odd parity have only recently emerged \cite{Jain:2021gwa,Jain:2021wyn,Buchbinder:2023coi}. Given the intricate nature of chiral and conformal anomalies, which are related to contact interactions, the coordinate approach becomes unwieldy, and the hierarchical character of the CWIs certainly becomes rather involved. Consequently, shifting to momentum space offers advantages due to its connection with ordinary off-shell scattering amplitudes.

\subsection{The $\langle TT J_5\rangle $ from parity-odd CFT}
Gravitational anomalies generated by spin 1/2 and spin 3/2 particles have been extensively studied in several works since the 70's, due to their connection with ordinary gauge theories \cite{Kimura:1969iwz,Delbourgo:1972xb}, supergravity  \cite{Nielsen:1978ex} and self-dual antisymmetric fields in string theory \cite{Alvarez-Gaume:1983ihn}, just to mention a few (see also \cite{Frolov:2022rod} for a recent study on the properties of chiral anomalies in the context of black holes).\\
The gravitational anomaly $R \tilde{R}$ can appear in different settings. 
The nonperturbative
approach developed in this paper is general and adaptable to many contexts, yet the anomaly's impact can vary from benign to dangerous, based on the circumstances. 
Consider, for example, a scenario involving a Dirac fermion interacting with gravity and a vector potential $V_\mu$. Kimura, Delbourgo and Salam were the first to compute the anomaly in this case, observed in the divergence of $J_5$ \cite{Kimura:1969iwz,Delbourgo:1972xb}. This specific anomaly poses no threat and may be of interest in phenomenology. For convenience, we can also introduce an axial-vector field $A_\mu$, which couples to $J_5$, but only as an external source since an anomalous gauge symmetry for $A_\mu$ would spoil unitarity and renormalizability.\\
Another instance involves a chiral model incorporating a Weyl fermion $\psi_{L/R}$ 
interacting with gravity and a gauge field. In this case, the anomaly emerges in the divergence of $J_{L/R}$, potentially endangering unitarity and renormalization, unless it is canceled \cite{Alvarez-Gaume:1983ihn}.  See \cite{Bonora:2023soh} for a detailed account on the types of chiral anomalies and their relation to diffeomorphism invariance. \\
In general, in perturbation theory, the evaluation of a chiral trace of Dirac matrices hinges on the choice of a specific regularization and the related treatment of the antisymmetric $\varepsilon$ tensor in the loop. In the case of the Breitenlohner-Maison-'t Hooft-Veltman scheme \cite{Breitenlohner:1977hr}, for example, the anomaly of parity-odd correlators is present only on the Ward identity of the chiral current, while the energy-momentum tensor and the vector currents 
are conserved. In other regularizations, one can potentially find a violation of the latter as well.\\
The correlator under scrutiny in this work is the $\langle TT J_5\rangle $, reinvestigated using CFT in momentum space (see \cite{Coriano:2020ees} for a review).
We will utilize a formalism developed for curved spacetime, from which the flat spacetime CWIs will be consistently derived in $d=4$, as constraints from special background metrics \cite{Coriano:2017mux}.
This correlator involves two stress-energy tensors and one parity-odd current, denoted as $J_5$. A study of this correlator was previously discussed in \cite{Erdmenger:1999xx} using coordinate space methods.
In the Standard Model, when $J_5$ is the non-Abelian $SU(2)$ gauge current or the hypercharge gauge current, this anomaly cancels out by summing over the chiral spectrum of each fermion generation. This feature is usually interpreted as an indication of the compatibility of the Standard Model when coupled to a gravitational background, providing an essential constraint on its possible extensions.
The correlator plays a crucial role in mediating anomalies of global currents associated with baryon $(B)$ and lepton $(L)$ numbers in the presence of gravity.\\
In condensed matter theory, correlators affected both by chiral and conformal anomalies, as well as by discrete anomalies, play an important role 
in the context of topological materials \cite{:2013kya,Ambrus:2019khr,Chernodub:2021nff,Arouca:2022psl,Tutschku:2020rjq}.
In our analysis, we demonstrate how investigating CFT in momentum space allows us to independently reproduce previous results found in coordinate space \cite{Erdmenger:1999xx}.\\
The solution is uniquely constructed by assuming the exchange of a single anomaly pole in the longitudinal sector of this correlator when we proceed with its sector decomposition. We will show that the momentum space solution, derived from the CFT constraints, is unique and depends on a single constant: the anomaly coefficient at the pole.\\
This result appears to be a common feature in correlation functions that are finite and affected by parity-odd anomalies, complementing our previous analysis of similar correlation functions such as the $\langle JJJ_5\rangle$ (or $\langle AVV\rangle$ chiral anomaly vertex) and the parity-odd $\langle TJJ\rangle$. \\
However, there are also some differences between the $\langle AVV\rangle $ and $\langle TTJ_5\rangle$ cases, which are affected by a chiral anomaly, and the $\langle TJJ\rangle_{odd} $, when this correlator is assumed to develop a parity-odd trace anomaly. In the $\langle AVV\rangle$ and the $\langle TTJ_5\rangle$, both the longitudinal and transverse sectors are nonzero and completely determined by the anomaly pole, initially introduced in the longitudinal sector as a solution of the anomaly constraint \cite{Coriano:2023hts}. Instead, in the $\langle TJJ\rangle_{odd} $, only the longitudinal sector survives after imposing the conformal constraints together with the (chiral) trace anomaly \cite{Coriano:2023cvf}, while the remaining sectors vanish.  \\

\subsection{Fermionic and Chern-Simons currents}
In free field theory, two realizations of currents have been discussed for this correlator: the bilinear (axial-vector) fermion current $J_{5 f}$ and the bilinear gauge-dependent Chern-Simons (CS) current $(J_5\equiv J_{CS})$ \cite{Dolgov:1988qx, Dolgov:1987yp}. 
We recall that in previous analysis it has been shown that both currents of the form 
\beq
J_{5 f}^{\lambda}=\bar{\psi}\gamma_5\gamma^\lambda \psi
\eeq
 or of the Chern-Simons form
\beq
J_{CS}^{\lambda} =\epsilon^{\lambda \mu\nu\rho} V_{\mu}\partial_{\nu} V_{\rho},
 \eeq 
 could be considered in a perturbative realization of the same correlator and generate a gravitational anomaly. 
Notice that this second version of the current can be incorporated into an ordinary partition function - in an ordinary Lagrangian realization by a path integral - only in the presence of a coupling to an axial-vector gauge field $(A_{\lambda})$ via an interaction of the form 
\beq
\mathcal{S}_{AVF}\equiv\int d^4 x \sqrt{g} A_\lambda J^{\lambda}_{CS}.
\eeq
The term, usually denoted as $A \, V\wedge F_V$, is the Abelian Chern-Simons form that allows to move the anomaly from one vertex to another in the usual $\langle AVV\rangle $ diagram. Details on these point can be found in \cite{Coriano:2005js,Armillis:2007tb}. 
Notice that both currents satisfy the parity-odd constraint given in \eqref{eq:anomaliachirale}. \\
$J_{CS}$ is responsible for mediating the gravitational chiral anomaly with spin-1 virtual photons in the loops, resulting in a difference between their two circular modes and inducing an optical helicity. This interaction is relevant in early universe cosmology and has an impact on the polarization of the Cosmic Microwave Background (CMB) \cite{Galaverni:2020xrq}. \\
In this case, the classical symmetry to be violated is the discrete duality invariance ($E\to B$, $B\to -E$) of the Maxwell equations in the vacuum (see \cite{Pasti:1995tn, Agullo:2018nfv}).
The $\langle TTJ_5\rangle$ correlator induces similar effects on gravitational waves \cite{delRio:2020cmv, delRio:2021bnl}. Spectral asymmetries induced by chiral anomalies, particularly the ordinary chiral anomaly (the $F\tilde{F}\sim E\cdot B$ term), have been investigated for their impact on the evolution of the primordial plasma, affecting the magneto-hydrodynamical (MHD) equations and the generation of cosmological magnetic fields \cite{Boyarsky:2011uy, Boyarsky:2015faa, Brandenburg:2017rcb}.\\
As previously mentioned, our method exclusively exploits the correlator's symmetries to identify its structure, which remains identical for a generic parity-odd $J_5$. In both cases $(J_{5f}$ and $J_{CS})$, the solution is entirely centered around the anomaly pole, serving as a pivot for the complete reconstruction of the corresponding correlators.\\
Both realizations of the $\langle TTJ_5\rangle$ correlator - using a $J_{CS}$ current or a $J_{5f}$ current - have been shown in \cite{Dolgov:1988qx, Dolgov:1987yp} to reduce to the exchange of an anomaly pole for on-shell gluons and photons for the unique form factors present in the diagrams.\\
In these works, the authors introduced a mass deformation of the propagators in the loops and showed the emergence of the pole as the mass was sent to zero. The method relies on the spectral density of the amplitude and it has been used also more recently in \cite{Giannotti:2008cv} and \cite{Coriano:2014gja}, in  studies of the parity-even $\langle TJJ\rangle$ and in supersymmetric variants.\\
We comment on this point in Section \ref{apolesumrule} and illustrate, by a simple computation, that the spectral densities of the only surviving form factors in the on-shell $\langle TT J_5\rangle$, with $J_5\equiv J_{5f}$ and $J_5\equiv J_{CS}$, satisfy two (mass-independent) sum rules.

\subsection{Organization of the work}
The outline of the paper is as follows.\\
In Section \ref{npappr} we review some of the feature of the perturbative approach. We examine the link between anomalies and the presence of poles in the expression of correlators in momentum space.\\
In Section \ref{rev}, we briefly comment on the methodology followed in the solution of the $4d$ CWIs in two previous analyses by us \cite{Coriano:2023hts, Coriano:2023cvf} involving anomalous parity-odd correlators, which may help clarify some of the technical points contained in this work.\\
Then, in Section \ref{ward}, we examine the constraints following from diffeomorphism, gauge and Weyl invariance. 
In particular, we express the conformal constraints on the $\langle TTJ_5\rangle$ as $4d$ differential equations first in coordinates and then in momentum space. 
The two following sections then discuss the general decomposition of the correlator, following the methods of \cite{Bzowski:2013sza}, extended to the parity-odd case, and the solution of the conformal constraints.
We present the general expression of the conformal $\langle TTJ_5\rangle$ correlator in momentum space. \\
In Section \ref{pertrealiz}, we perform a perturbative analysis of the correlator and in Section \ref{matchpert} we verify that the conformal solution and the perturbative one coincide.\\
Then, in Section \ref{apolesumrule} we show that, similarly to previous dispersive analysis of the anomalous form factors for the $\langle TJJ\rangle $ and $\langle AVV\rangle$ diagrams,
the spectral density of the anomalous form factor of the $\langle TTJ_5\rangle$ satisfies a sum rule.\\
We summarize our findings in Section \ref{sumresultsec} and discuss the non-renormalizability of the $\langle AVV\rangle$ and $\langle TTJ_5\rangle$ in Section \ref{commnonrinormsec}.
We leave to the Appendix a discussion of some technical points concerning 3K integrals, the use of the Schouten relations and the identities we have used to identify the correlator by the perturbative and non-perturbative methods.\\
 It is worth noting that the analysis of the conformal constraints and the reconstruction method of \cite{Bzowski:2013sza}, here implemented in the parity-odd case, allows us to express the correlator in terms of a minimal number of form factors. For example, in the case of the parity-even $\langle TJJ\rangle$, the reduction in their numbers has been from thirteen down to three \cite{Bzowski:2013sza}\cite{Coriano:2018zdo} by the inclusion of the conformal constraints and a special choice of the parametrization of the tensor structures symmetric in the external momenta. A similar analysis has been performed for 4-point functions in the parity-even $\langle TTJJ\rangle$ \cite{Coriano:2022jkn} and $\langle TOOO\rangle$ \cite{Coriano:2019nkw}, where $O$ are identical conformal primary scalars.

\section{The nonperturbative approach}\label{npappr}
We will investigate the structure of the \textit{full} (uncontracted) vertices, in contrast to the majority of previous literature that has focused solely on computing the anomaly, namely the WI.\\
As demonstrated in previous works through perturbative analyses of correlators such as the $\langle JJ J_5\rangle$ and the parity-even $\langle TJJ\rangle$  \cite{Giannotti:2008cv,Armillis:2009pq}, or the superconformal anomaly multiplet \cite{Coriano:2014gja}, the anomaly in momentum space is associated with the exchange of an anomaly pole \cite{2009PhLB..682..322A}. While the coordinate space approach is valuable, it has limitations in revealing the underlying dynamical source of the anomaly. This aspect becomes considerably clearer in momentum space when conducting a dispersive analysis of the anomaly form factor.\\
In the case of the $\langle TTJ_5\rangle$, previous analyses, conducted for $J_{CS}$, have identified the presence of such pole in the correlator \cite{Dolgov:1988qx}. This analysis was based on perturbation theory for on-shell gravitons. We will provide further comments on these previous findings in a following section.\\
It has been established, in a somewhat general context, that various types of anomalies (chiral, conformal, supersymmetric, etc.) are invariably characterized by the presence of a form factor with a pole, which multiplies a tensor structure responsible for generating the anomaly. The nonlocality of the anomaly is, therefore, summarized, at least in the context of flat spacetime, by an effective action of the form \cite{Giannotti:2008cv}\cite{Armillis:2009pq}.

 \beq
 \mathcal{S}_{JJJ_5} =\int d^4 x d^4 y \partial \cdot A \,\Box^{-1}(x,y) F_V\tilde F_V(y) +\ldots
 \label{chir}
 \eeq 
 for an anomalous $\langle J_5 JJ\rangle$ (axial-vector/vector/vector) vertex, with $A^\mu$ denoting an external axial vector and $V_{\mu}$ a vector gauge field. A similar nonlocal action 
 \beq
 \mathcal{S}_{TJJ} \int d^4 x d^4 y R^{(1)}(x) \,\Box^{-1}(x,y) F_V^{} F_{V} (y)+\ldots
 \label{conf}
 \eeq  
 can be written down for the parity-even $\langle TJJ\rangle$ vertex, with one stress-energy tensor $T$ and two conserved vector currents $J$ \cite{Giannotti:2008cv,Armillis:2010qk}. $R^{(1)}$ is the linearized Ricci scalar.  The ellipsies in the expressions above indicate terms which do not contribute to the anomaly, either chiral \eqref{chir} or conformal \eqref{conf}. 
 If we identify the $ \langle AVV \rangle$ vertex in momentum space as
 \be
  \langle J^{\mu_1}(p_1)J^{\mu_2}(p_2)J_5^{\mu_3}(p_3)\rangle=\int d^4 x_1\, d^4 x_2\, d^4 x_3\, e^{-i p_1\cdot x_1 -i p_2\cdot x_2-i p_3\cdot x_3} \langle J^{\mu_1} (x_1)J^{\mu_2}(x_2) J_5^{\mu_3}(x_3) \rangle
 \ee
 the contribution isolated in \eqref{chir} is obtained from the solution of the constraint
\begin{equation}
	p_{3 \mu_3}\left\langle J^{\mu_1}\left(p_1\right) J^{\mu_2}\left(p_2\right) J_5^{\mu_3}\left(p_3\right)\right\rangle=-8 \,a_1\, i\, \varepsilon^{ \mu_1 \mu_2\alpha \beta}p_{1\alpha}p_{2\beta}\equiv -8 \,a_1\, i\, \varepsilon^{ \mu_1 \mu_2 p_1p_2}
\end{equation}
 in the form 
 \beq
\langle J^{\mu_1}(p_1)J^{\mu_2}(p_2)J_5^{\mu_3}(p_3)\rangle= -8 \,a_1 i\, \frac{p_3^{\mu_3}}{p_3^2}\epsilon^{\mu_1\mu_2p_1p_2} + \ldots
  \eeq
 with the ellipsis referring to to the transverse components. The $1/p_3^2$ contribution is the anomaly pole.  \\
 In a similar fashion, in the $\langle TJJ\rangle$, the momentum space analysis reveals that the correlator is decomposed in the form \cite{Giannotti:2008cv,Armillis:2009pq}
 \beq
\langle T^{\mu_1\nu_1}(p_1)J^{\mu_2}(p_2)J^{\mu_3}(p_3)\rangle=\frac{b_4}{3 p_1^2}\left(\delta^{\mu_1\nu_1}p_1^2 -p_1^{\mu_1} p_1^{\nu_1} \right) u^{\mu_2\mu_3}(p_2,p_3)
 \eeq
where $b_4$ is the anomalous coefficient in eq$.$ \eqref{ann} and
 \beq
u^{\mu_2\mu_3}(p_2,p_3) \equiv (p_2\cdot p_3) g^{\mu_2\mu_3} - p_2^{\mu_3}p_3^{\mu_2}\,,\\ 
 \eeq
 is given by the Fourier transform of the anomaly functional ($F\tilde F$) differentiated with respect to the external gauge field
 \beq
 u^{\mu_2\mu_3}(p_2,p_3) = -\frac{1}{4}\int\,d^4x_2\,\int\,d^4x_3\ e^{-ip_2\cdot x_2 - i p_3\cdot x_3}\ 
\frac{\delta^2 \{{F}_{V\mu\nu}F_V^{\mu\nu}(0)\}} {\delta V_{\mu_2}(x_2)\delta V_{\mu_3}(x_3)}. 
 \eeq
 It is evident that, based on this information, one anticipates that each anomalous vertex should exhibit a nonlocal interaction in flat spacetime, mirroring the two aforementioned instances. What presents a greater challenge is the determination of whether the inclusion of a pole as a solution to the anomalous WI, in conjunction with the conformal constraints, suffices to ascertain the complete expression of the correlators. In other words, concerning the parity-odd correlators $\langle TJJ\rangle$, $\langle AVV\rangle$, and $\langle TTJ_5\rangle$, it is conceivable that the 3-point functions can be wholly reconstructed from the residue at the pole, which is defined by the anomaly coefficient.\\
It is worth noting that, within our methodology, the pole is not directly associated with the current but with the entire correlator itself. Hence, in the context of the $\langle TTJ_5\rangle$ correlator under investigation in this study, the specific form of the parity-odd operator current that satisfies the anomaly constraint does not play an essential role.\\
 
\section{Parity-odd terms}  
\label{rev}
A common track of parity-odd correlation functions is the presence of anomaly constraints, either in the form of chiral trace anomalies, or just of chiral anomalies, inducing CP violation.  In the case of chirally-odd trace anomalies, which have been at the center of several recent debates,  
the anomaly functional, constraining the form of the correlator, is related to CP violating terms beside the usual parity-even ones in the form 
\begin{equation}
\label{ann}
	g_{\mu\nu}\langle T^{\mu \nu}\rangle= b_1\, {E}_4+b_2 \, C^{\mu \nu \rho \sigma} C_{\mu \nu \rho \sigma}+b_3\, \nabla^2 R+b_4\, F^{\mu\nu}F_{\mu\nu}+f_1\, \varepsilon^{\mu \nu \rho \sigma} R_{\alpha \beta \mu \nu} R_{\,\,\,\,\, \rho \sigma}^{\alpha \beta}+f_2\,\varepsilon^{\mu\nu\rho\sigma}F_{\mu\nu}F_{\rho\sigma}.
\end{equation}
where $F_{\mu \nu}$ can be either an axial or vector gauge field.
Parity-odd terms corresponding to the constant $f_1$ and $f_2$ have been conjectured long ago 
by Capper and Duff \cite{Capper:1974ic,Duff:2020dqb} on dimensional grounds and by requiring covariance. From a cohomological point of view, such term are consistent and cannot be discarded. In previous work we have shown that the CWIs severely constrain the structure of parity-odd correlators \cite{Coriano:2023cvf}.  For example, a correlator such as the (rank-4)
$\langle TJJ \rangle_{odd}$, if we resort to a longitudinal/transverse-traceless/trace decomposition of its tensorial structure, is constrained only to contain a nonvanishing trace sector, with the remaining sectors being zero. We will illustrate a similar decomposition in the case of the $\langle TT J_5\rangle$, that we are going to investigate in the next sections.   \\
A similar analysis has been performed in the case of the $\langle AVV\rangle$  diagram \cite{Coriano:2023hts}, decomposed into a transverse and a longitudinal sectors. We have shown that the entire chiral anomaly vertex  can be derived from CFT by solving the anomaly constraints, with the anomaly attributed to the axial-vector current 
\beq
\partial_\a \langle J_5^\a \rangle= a_1\, \varepsilon^{\mu \nu \rho \sigma} F_{\mu \nu} F_{\rho \sigma} 
\eeq
 The reconstruction of the correlator is performed using as a pivot the anomaly pole present in the decomposition of this correlator in its anomaly sector. In the case of the $\langle JJJ_5\rangle $ or of the $\langle J_5 J_5 J_5\rangle $, we satisfy the anomaly constraint with one or three anomaly poles respectively, and all the remaining sectors are fixed by this choice. The general character of this reconstruction procedure, based on the inclusion of the anomaly constraint, will be proven also in the case of the $\langle TT J_5\rangle$. In all these three cases, our analysis shows that the anomaly phenomenon, at least in the parity odd case, is entirely associated with the presence of an anomaly pole. Once the coefficient in front of the anomaly is determined, then the entire correlator is fixed, if we impose the conformal symmetry. In the $\langle TTJ_5\rangle$, if we include in the background both a gauge field and a general metric background, the anomalous WI that we will be using for the definition of the correlator in CFT is given by the equation 

\begin{equation} \label{eq:anomaliachirale}
	\nabla_\mu\langle J_5^\mu\rangle=
	a_1\, \varepsilon^{\mu \nu \rho \sigma}F_{\mu\nu}F_{\rho\sigma}+ a_2 \, \varepsilon^{\mu \nu \rho \sigma}R^{\alpha\beta}_{\hspace{0.3cm} \mu \nu} R_{\alpha\beta \rho \sigma},
\end{equation}
that defines a boundary condition for the CWIs.\\
These results can also be extended to the non-Abelian case. For a general chiral current $J^\mu_i$, we can write 
\begin{equation}
	\nabla_\mu \langle J_i^\mu \rangle= a_1\,D_{ijk} \, \varepsilon^{\mu \nu \rho \sigma} F^j_{\mu \nu} F^k_{\rho \sigma} + a_2 \,D_i \,\varepsilon^{\mu \nu \rho \sigma}R^{\alpha\beta}_{\hspace{0.3cm} \mu \nu} R_{\alpha\beta \rho \sigma},
\end{equation}
where we have introduced the anomaly tensors
\begin{equation}
	D_{ijk} =\frac{1}{2} \, \text{Tr}[\{T_i,\,T_j\}\,T_k],\quad \qquad D_i=\text{Tr}[T_i]. 
\end{equation}
constructed with the non-Abelian generators of the theory. In the case of the $D_i$'s, for example, in the Standard Model, where the symmetry is $SU(3)\times SU(2)\times U(1)_Y$, only the hypercharge $(U(1)_Y)$ contribution $\langle TTJ_Y\rangle$ is taken onto account, since the $SU(2)$ and $SU(3)$ generators are traceless.\\
 Both the chiral $(F\tilde{F})$ and gravitational $(R\tilde{R})$ anomalies cancel once we sum over each generation of chiral fermions \cite{Weinberg:1996kr}. 
The cancellation of the gravitational anomaly in the Standard Model can be interpreted in two possible ways. On one hand, it shows the consistency of the coupling of the Standard Model to gravity, since the gauge currents are conserved in a gravitational background. On the other hand, the stress-energy tensor is just another operator of the Standard Model and the conservation  of the currents is required for the analysis of the mixing of such operator with the gauge currents at perturbative level.

\section{Ward identities}
\label{ward}
The correlator is constrained by diffeomorphism invariance, gauge invariance and the CWIs, derived from Weyl invariance. For this purpose, it is convenient to obtain the identities directly from the functional integral, by requiring the invariance of the effective action $\cS$ in the Euclidean space
\be
e^{-\cS[g]} \equiv \int [d \Phi]\, e^{- {{S}_{0}}[\Phi,\, g]}
\label{Sexact}
\ee
under the corresponding symmetries. The integration runs over all the matter/radiation fields, here denoted as $\Phi$. The equations can also define the constraints for any non-Lagrangian CFT. \\
We couple the external gauge field source $A_\mu$ to the current $J_5$.
We define the quantum averages of the energy-momentum tensor and of the current in terms of the generating functional of the theory $\cS$
\begin{equation}
	\left\langle T^{\mu \n}(x)\right\rangle=\left.\frac{2}{\sqrt{-g(x)}} \frac{\delta \cS}{\d g_{\mu \nu}(x)}\right|_{g=\delta}, \quad \quad\left\langle J_5^{\mu}(x)\right\rangle=\left.\frac{1}{\sqrt{-g(x)}} \frac{\delta \cS}{\d A_{\mu}(x)}\right|_{A=0}
\end{equation}
from which we derive the correlator of interest

\begin{equation}
	\left\langle T^{\mu_1 \nu_1}\left({x}_{1}\right) T^{\mu_2 \nu_2}\left({x}_{2}\right) J_5^{\mu_3 }\left({x_3}\right)\right\rangle\equiv 
	\left. \frac{2^2}{\sqrt{-g(x_1)}\sqrt{-g(x_2)}\sqrt{-g(x_3)}}\frac{\delta^3 \cS}{\d g_{\mu_1 \nu_1}(x_1)\d g_{\mu_2 \nu_2}(x_2)\d A_{\mu_3}(x_3)}\right|_{\scriptsize{\begin{aligned}
				{g=\delta;}\\{A=0;}
	\end{aligned}}}.
\end{equation}
\begin{itemize}
\item{\bf Diffeomorphism invariance}
\end{itemize}
We start from diffeomorphism invariance. \\
Under a diffeomorphism the fields transform with a Lie derivative
\begin{equation}
	\begin{aligned}
		\delta g_{\mu \n} &=\nabla_{\mu} \xi_{\nu}+\nabla_{\nu} \xi_{\mu}\\
		\delta A_{\mu} &=\xi^{\n} \nabla_{\n} A_{\mu}+\nabla_{\mu} \xi^{\n} A_{\n}.
	\end{aligned}
\end{equation}
The WIs follow from the requirement that the generating functional $\cS$ is invariant under these transformations 
\begin{equation}
	\begin{aligned}
		0=\delta_{\xi} \cS &=\int d^{d} x\left[ \left(\nabla_{\mu} \xi_{\n}+\nabla_{\n} \xi_{\mu}\right)  \frac{\delta}{\delta g_{\mu \nu}}+\left(\xi^{\n} \nabla_{\nu} A_{\mu}+\nabla_{\mu} \xi^{\n} A_{\n}\right) \frac{\delta}{\delta A_{\mu}}\right] \cS \\
				&=\int d^{d} x \sqrt{-g} \xi^{\n}\left[-\nabla^{\mu}\left\langle T_{\mu \nu}(x)\right\rangle+\nabla_{\nu} A_{\mu}\left\langle J_5^{\mu }(x)\right\rangle-\nabla_{\mu}\left(A_{\n}\left\langle J_5^{\mu }(x)\right\rangle\right) \right]
	\end{aligned}
\end{equation}
obtaining
\begin{equation} \label{eq:WIdiff}
	\nabla^{\mu}\left\langle T_{\mu \nu}\right\rangle-{F_A}_{ \n \m} \left\langle J_5^{\mu }\right\rangle+A_{\n}\nabla_{\mu}\left\langle J_5^{\mu }\right\rangle=0.
\end{equation}
In order to find a WI for the correlator $\langle TTJ_5 \rangle$ we need to apply to this equation the functional derivatives $\frac{\delta }{\delta g \, \delta A}$. Due to the vanishing of all 1-point and 2-point functions in $4d$ in the limit $g_{\mu \nu}\rightarrow \delta_{\mu \nu}$ and $A_\mu\rightarrow 0$, only the first term in the eq. \eqref{eq:WIdiff} survives.
Going to momentum space, this procedure generates the equation
\begin{equation}
	0=p_{i\mu_i} \left\langle T^{\mu_1 \nu_1}\left({p}_{1}\right) T^{\mu_2 \nu_2}\left({p}_{2}\right) J_5^{\mu_3 }\left({p_3}\right)\right\rangle, \qquad\qquad\qquad i=1,2.
\end{equation}

\begin{itemize}
\item{\bf Gauge invariance}
\end{itemize}
We proceed in a similar manner for  gauge invariance.
The action of a gauge transformation, with parameter $\a(x)$ on the fields, gives the infinitesimal variations
\begin{equation}
	\begin{aligned}
		\delta g_{\mu \nu} &=0, \qquad\qquad 
		\delta A_{\mu} =\partial_{\mu} \alpha.
	\end{aligned}
\end{equation}
The requirement that the generating functional $\cS$ is invariant under these transformations leads to the conservation of the current $J_5^\a$. If we allow an anomaly from the path integral measure we obtain the relation
\begin{equation} \label{eq:curcons}
	\nabla_\a \langle J_5^\a \rangle= a_1\, \varepsilon^{\mu \nu \rho \sigma} F_{\mu \nu} F_{\rho \sigma} +a_2\, \varepsilon^{\mu \nu \rho \sigma} R_{\a \beta \mu \nu} R_{\quad \rho \sigma}^{\a \beta},
\end{equation}
We now apply two functional derivatives with respect to the metric to such equation, and perform the limit $g_{\mu \nu}\rightarrow \delta_{\mu \nu}$ and $A_\mu\rightarrow 0$ as above. After going to the momentum space, we derive the relation
\small
\begin{equation}\label{eq:wigauge3}
	p_{3\mu_3}\braket{T^{\mu_1\nu_1}(p_1)T^{\mu_2\nu_2}(p_2)J_5^{\mu_3}(p_3)}= 4\, i \, a_2 \, (p_1 \cdot p_2) \left\{ \left[\varepsilon^{\nu_1 \nu_2 p_1 p_2}\left(g^{\mu_1 \mu_2}- \frac{p_1^{\mu_2} p_2^{\mu_1}}{p_1 \cdot p_2}\right) +\left( \mu_1 \leftrightarrow \nu_1 \right) \right] +\left( \mu_2 \leftrightarrow \nu_2 \right) \right\}.
\end{equation}
\normalsize
This constraint will be satisfied by the inclusion of an anomaly pole in the correlator.

\begin{itemize}
\item{\bf Weyl invariance}
\end{itemize}
 By requiring the invariance of the partition function under the conformal transformations in flat space, we identify five extra constraints, beside those related to diffeomorphism invariance, corresponding to the special conformal transformations and the dilatations. Taken together, diffeomorphism plus Weyl invariance in curved spacetime determine in any local free-falling frame associated with the metric $g_{\mu\nu}$, the symmetry constraints of the conformal group on the correlator. \\
 The derivation of these constraints, from a curved spacetime perspective, requires a background metric that allows conformal Killing vectors. This imposes a formal restriction on the class of metric backgrounds with respect to which the functional variations are performed. In other words, this approach allows to patch together the constraints in each local frame.  If the classical action is Weyl invariant, a metric that allows conformal Killing vectors  requires that $\sigma(x)$ is at most quadratic in the local coordinates, and  is expressed in terms of 15 parameters, which are the parameter of the conformal group.   
 We try to clarify this point, that has been discussed in \cite{Coriano:2017mux} in the case of the $\langle TTT\rangle$ correlator, extending it to the $\langle TTJ_5\rangle $, with its specific expression of the anomaly. \\
 The action of a Weyl transformation with a dilaton $\sigma(x)$ on the fields, acting as a parameter is
\begin{equation}\label{eq:trasfweyll}
	\begin{aligned}
		\delta g_{\mu \nu} &=2 g_{\mu \nu} \sigma, \\
		\delta A_{\mu} &=0.
	\end{aligned}
\end{equation}
If the metric is selected in such a way to allow conformal Killing vectors, then the diffeomorphisms 
$x^\mu\to x'^{\mu}=x^{\mu} + K^\mu(x)$ induce a simple rescaling of the infinitesimal distance, under a local rescaling with $\sigma(x)$
\begin{equation}
(ds')^2 = e^{2 \sigma(x)}(ds)^2. 
\end{equation}
This require that $\sigma(x)$ and the same vectors are related 
\beq
\label{cke}
  \nabla_\mu K_\nu+ \nabla_\nu K_\mu= 2 \sigma \delta_{\mu\nu}\qquad \sigma=\frac{1}{d}\nabla\cdot K.
\end{equation}
At this stage, from the action we define the conformal currents $J_K$, with their quantum averages given by
 \beq
\langle J_K^{\mu }\rangle \equiv K_{\nu}\langle T^{\mu\nu}\rangle
\eeq
that differentiated give 
\beq
\nabla_{\mu} \langle J_K^{\mu }\rangle = \frac{1}{2}\left(\nabla_\mu K_{\nu} + \nabla_\nu K_{\mu}\right) \langle T^{\mu\nu}\rangle +K_{\nu}\nabla_\mu \langle T^{\mu\nu}\rangle. 
\eeq
At this stage we can use both the property of the background metric \eqref{cke} and resort to the definition of the stress energy tensor to rewrite the equation into the form 
\beq
\label{add}
\sqrt{g} \nabla_\mu \langle J^{\mu}_K\rangle=\sigma \frac{\delta}{\delta\sigma}\langle {S}_{0}\rangle
+2 K_{\nu}\nabla_{\mu}\langle \frac{\delta}{\delta g_{\mu\nu}}{S}_{0}\rangle.
  \eeq
  where
  \begin{equation}
  	\frac{\delta}{\delta \sigma}\equiv 2 g_{\mu\nu}\frac{\delta}{\delta g_{\mu \nu}}
  \end{equation}
The conservation of the conformal currents 
\beq
\nabla_\mu \langle J^{\mu}_K\rangle=0,
\eeq
requires that 
\beq
\label{add1}
\frac{\delta}{\delta\sigma} {S}_{0}=0,
\eeq
plus the ordinary diffeomorphism invariance, that guarantees the vanishing of the second addend in \eqref{add}
\beq
\nabla_{\mu}\langle \frac{\delta}{\delta g_{\mu\nu}}{S}_{0}\rangle=0.
\eeq
 At quantum level  \eqref{add1} may be affected by a conformal anomaly  
with 
\beqa
\frac{\delta}{\delta\sigma}\langle {S}_{0}\rangle &=&\sqrt{g}\langle T^{\mu}_{\mu}\rangle\nn\\
&=&\sqrt{g}\mathcal{A},
\eeqa
where  $\mathcal{A}$ is the anomaly functional. The conservation of $J_K$ is violated in the case of such conformal
anomaly. \\
The invariance of the generating functional $\cS$ under the transformations \eqref{eq:trasfweyll} leads to the tracelessness of the energy-momentum tensor. However, in general we need to consider anomalous terms coming from the path integral measure
\begin{equation}
\label{AN1}
	\mathcal{A}=g_{\mu\nu}\langle T^{\mu \nu}\rangle= b_1\, {E}_4+b_2 \, C^{\mu \nu \rho \sigma} C_{\mu \nu \rho \sigma}+b_3\, \nabla^2 R+b_4\, F^{\mu\nu}F_{\mu\nu}+f_1\, \varepsilon^{\mu \nu \rho \sigma} R_{\alpha \beta \mu \nu} R_{\,\,\,\,\, \rho \sigma}^{\alpha \beta}+f_2\,\varepsilon^{\mu\nu\rho\sigma}F_{\mu\nu}F_{\rho\sigma}.
\end{equation}
After applying the functional derivatives $\frac{\d}{\d g} \frac{\d}{\d A}$ and performing the limit $g_{\mu \nu}\rightarrow \delta_{\mu \nu}$ and $A_\mu\rightarrow 0$, the anomalous terms do not survive.
Going to momentum space, we obtain the constraint
\begin{equation}
	g_{\mu_i \nu_i} \langle T^{\mu_1 \nu_1}({p}_1) T^{\mu_2 \nu_2}({p}_2) J_5^{\mu_3}  ({p}_3) \rangle=0,\qquad \qquad i=1,2
\end{equation}
which is a nonanomalous trace WI. \\
We now consider the conformal transformations.
The conformal Killing vectors for the dilatations are
\be
\label{CKVD}
K_{\m}^{(D)}(x) \equiv  x_\m\,,\qquad\qquad \pa\cdot K^{(D)} = d,
\ee
while for the special conformal transformation they are given by
\be
\label{CKVS}
K^{(S)\, \ka}_{\ \ \ \m}(x) \equiv 2x^{\ka}x_{\m} - x^2 \del^{\ka}_{\ \,\m}\,,
\qquad\qquad \pa\cdot K^{(S)\,\ka}(x) = (2d)\,x^{\ka}\,,\qquad\qquad\ka =1, \dots, d. 
\ee
As discussed in \cite{Coriano:2017mux}, the derivation of the CWIs requires a rank-4 correlator and can be directly formulated in the local frame. In this case, we focus on the 4-point function $\langle TTT J_5\rangle$ and consider the divergence condition
\begin{equation}
	\begin{aligned}
		0=&\int  dx \, \partial_\mu^{(x)} \bigg[K_\nu(x) \langle T^{\mu\nu}(x)T^{\mu_1\nu_1}(x_1)T^{\mu_2\nu_2}(x_2)J_5^{\mu_3}(x_3)\rangle \bigg]=\\&
		\int dx \,\left( \partial_\mu K_\nu\right) \langle T^{\mu\nu}(x)T^{\mu_1\nu_1}(x_1)T^{\mu_2\nu_2}(x_2)J_5^{\mu_3}(x_3)\rangle+
		K_\nu \partial_\mu \langle T^{\mu\nu}(x)T^{\mu_1\nu_1}(x_1)T^{\mu_2\nu_2}(x_2)J_5^{\mu_3}(x_3)\rangle.
	\end{aligned}
\end{equation}
Recalling the conformal Killing vector equation \eqref{cke}, we can then derive the equation
\begin{equation}\label{eq:ckvtjjt}
	\begin{aligned}
		0=
		\int dx \,\left( \frac{\partial\cdot K}{d}\right)&\eta_{\mu\nu} \langle T^{\mu\nu}(x)T^{\mu_1\nu_1}(x_1)T^{\mu_2\nu_2}(x_2)J_5^{\mu_3}(x_3)\rangle\\&
		+
		K_\nu \partial_\mu \langle T^{\mu\nu}(x)T^{\mu_1\nu_1}(x_1)T^{\mu_2\nu_2}(x_2)J_5^{\mu_3}(x_3)\rangle.
	\end{aligned}
\end{equation}
On the right-hand side of the last equation we have the trace and the divergence of a four-point correlator function. We can use the anomalous trace equation and the conservation of the energy-momentum
tensor in order to rewrite such terms. We will show this in the following.
We first focus on the dilatations. The Killing vectors in this case are given by (23). 
The invariance under diffeomorphism of the partition function \eqref{eq:WIdiff}, differentiated and in the flat limit gives
\begin{equation}
	\begin{aligned}
		0=\partial_\mu \langle& T^{\mu\nu}(x)T^{\mu_1\nu_1}(x_1)T^{\mu_2\nu_2}(x_2)J_5^{\mu_3}(x_3)\rangle+\\&
		\bigg[ \left(\partial_\mu \delta_{xx_1}\right)\delta^{\nu\mu_1}\delta_{\lambda}^{\nu_1}+
		\left(\partial_\mu \delta_{xx_1}\right)\delta^{\nu\nu_1}\delta_{\lambda}^{\mu_1}-
		\left(\partial^\nu \delta_{xx_1}\right)\delta^{\mu_1}_\mu \delta^{\nu_1}_\lambda
		\bigg] \langle T^{\lambda\mu}(x)T^{\mu_2\nu_2}(x_2)J_5^{\mu_3}(x_3)\rangle +\\&
		\bigg[ \left(\partial_\mu \delta_{xx_2}\right)\delta^{\nu\mu_2}\delta_{\lambda}^{\nu_2}+
		\left(\partial_\mu \delta_{xx_2}\right)\delta^{\nu\nu_2}\delta_{\lambda}^{\mu_2}-
		\left(\partial^\nu \delta_{xx_2}\right)\delta^{\mu_2}_\mu \delta^{\nu_2}_\lambda
		\bigg] \langle T^{\lambda\mu}(x)T^{\mu_1\nu_1}(x_1)J_5^{\mu_3}(x_3)\rangle +\\&
		\bigg[-\left(\partial_\nu \delta_{xx_3}\right)\delta^{\mu_3}_\mu+\left(\partial_{\mu}\delta_{xx_3}\right)\delta_\nu^{\mu_3}
		+\delta^{\mu_3}_\nu \delta_{xx_3}\partial_\mu 
		\bigg]\langle T^{\mu_1\nu_1}(x_1)T^{\mu_2\nu_2}(x_2)J_5^{\mu}(x)\rangle 
	\end{aligned}
\end{equation}
where $\delta_{xy}$ is the dirac delta function $\delta^4 (x -y)$ and all the derivatives are with respect to the $x$ variable. Similarly, we differentiate the anomalous trace equation \eqref{eq:WIdiff} obtaining in the flat limit
\begin{equation}
	\begin{aligned}
		\delta_{\mu \nu} \langle& T^{\mu\nu}(x)T^{\mu_1\nu_1}(x_1)T^{\mu_2\nu_2}(x_2)J_5^{\mu_3}(x_3)\rangle=
		-2\bigg( \delta_{xx_1}  +\delta_{xx_2}  \bigg)\langle T^{\mu_1\nu_1}(x_1)T^{\mu_2\nu_2}(x_2)J_5^{\mu_3}(x_3)\rangle.
	\end{aligned}
\end{equation}
Inserting the equations (23), (30) and (32) into (28) and integrating by parts we obtain the ordinary dilatation Ward identity (WI)

\begin{equation}
	0=\bigg[ \sum_{i} x_i^\mu \, \partial_\mu^{(x_i)}+3d-1
	\bigg]\langle T^{\mu_1\nu_1}(x_1)T^{\mu_2\nu_2}(x_2)J_5^{\mu_3}(x_3)\rangle,
\end{equation}
 that in momentum space takes the form
\begin{align}
\label{eqD}
	\left(\sum_{i=1}^3\Delta_i-2d-\sum_{i=1}^2\,p_i^\mu\frac{\partial}{\partial p_i^\mu}\right)\braket{T^{\mu_1\nu_1}(p_1)T^{\mu_2\nu_2}(p_2)J_5^{\mu_3} (p_3)}=0.
\end{align}
If, instead, we consider the special conformal transformations, the Killing vectors are given in (24). Proceeding in a similar manner we arrive at the expression

\begin{align}
	&0=\sum_{i=1}^3\left[2 x_i^\kappa\left(\Delta_i+x_i^\alpha \frac{\partial}{\partial x_i^\alpha}\right)-x_i^2 \delta^{\kappa \alpha} \frac{\partial}{\partial x_i^\alpha}\right]
	\braket{T^{\mu_1\nu_1}(x_1)T^{\mu_2\nu_2}(x_2)J_5^{\mu_3}({x}_3)}\nonumber\\&+
	2\bigg[\delta^{\kappa \mu_1}{x_1}_\alpha-\delta^\kappa_\alpha x_1^{\mu_1}
	\bigg]
	\braket{T^{\alpha\nu_1}(x_1)T^{\mu_2\nu_2}(x_2)J_5^{\mu_3}({x}_3)}
	+
	2\bigg[\delta^{\kappa \nu_1}{x_1}_\alpha-\delta^\kappa_\alpha x_1^{\nu_1}
	\bigg]
	\braket{T^{\mu_1\alpha}(x_1)T^{\mu_2\nu_2}(x_2)J_5^{\mu_3}({x}_3)}
	\nonumber\\&+
	2\bigg[\delta^{\kappa \mu_2}{x_2}_\alpha-\delta^\kappa_\alpha x_2^{\mu_2}
	\bigg]
	\braket{T^{\mu_1\nu_1}(x_1)T^{\alpha\nu_2}(x_2)J_5^{\mu_3}({x}_3)}
	+
	2\bigg[\delta^{\kappa \nu_2}{x_2}_\alpha-\delta^\kappa_\alpha x_2^{\nu_2}
	\bigg]
	\braket{T^{\mu_1\nu_1}(x_1)T^{\mu_2\alpha}(x_2)J_5^{\mu_3}({x}_3)}\nn \\&  +
	2\bigg[\delta^{\kappa \mu_3}{x_3}_\alpha-\delta^\kappa_\alpha x_3^{\mu_3}
	\bigg]
	\braket{T^{\mu_1\nu_1}(x_1)T^{\mu_2\nu_2}(x_2)J_5^{\alpha}({x}_3)},
\end{align}
that in momentum space takes the form
\begin{equation}
	\begin{aligned}
		0=&\, \mathcal{K}^\kappa\left\langle T^{\mu_1 \nu_1}\left(p_1\right) T^{\mu_2 \nu_2}\left(p_2\right) J_5^{\mu_3 }\left(p_3\right)\right\rangle \\
		&=\sum_{j=1}^2\left(2\left(\Delta_j-d\right) \frac{\partial}{\partial p_{j \kappa}}-2 p_j^\alpha \frac{\partial}{\partial p_j^\alpha} \frac{\partial}{\partial p_{j \kappa}}+\left(p_j\right)^\kappa \frac{\partial}{\partial p_j^\alpha} \frac{\partial}{\partial p_{j \alpha}}\right)\left\langle T^{\mu_1 \nu_1}\left(p_1\right) T^{\mu_2 \nu_2}\left(p_2\right) J_5^{\mu_3}\left(p_3\right)\right\rangle \\
		&\qquad+4\left(\delta^{\kappa\left(\mu_1\right.} \frac{\partial}{\partial p_1^{\alpha_1}}-\delta_{\alpha_1}^\kappa \delta_\lambda^{\left(\mu_1\right.} \frac{\partial}{\partial p_{1 \lambda}}\right)\left\langle T^{\left.\nu_1\right) \alpha_1}\left(p_1\right) T^{\mu_2 \nu_2}\left(p_2\right) J_5^{\mu_3 }\left(p_3\right)\right\rangle \\
		&\qquad+4\left(\delta^{\kappa\left(\mu_2\right.} \frac{\partial}{\partial p_2^{\alpha_2}}-\delta_{\alpha_2}^\kappa \delta_\lambda^{\left(\mu_2\right.} \frac{\partial}{\partial p_{2 \lambda}}\right)\left\langle T^{\left.\nu_2\right) \alpha_2}\left(p_2\right) T^{\mu_1 \nu_1}\left(p_1\right) J_5^{\mu_3 }\left(p_3\right)\right\rangle .
	\end{aligned}
\end{equation}
At this stage we are ready to proceed with the decomposition of the correlator into all of its sectors and derive the scalar equations for its reconstruction \cite{Bzowski:2013sza}. We will first proceed with a parametrizaton of the form factors and tensors structures of the transverse-traceless sector. We introduce a form factor in the longitudinal part, in the form of an anomaly pole, and proceed with a complete determination of the entire correlation function by solving the equations of all the remaining sectors. We follow the steps introduced in \cite{Bzowski:2013sza}, extended to the parity-odd case, and split the equations into primary and secondary CWIs. 
The solution, as we are going to show, will coincide with the perturbative one and will depend on a single constant, the coefficient of the anomaly. The off-shell parametrization of the vertex  that results from this construction is quite economical, and is expressed in terms of only two form factors in the transverse traceless sector, plus the anomaly form factor that takes the form of a $1/p_3^2$ anomaly pole. 
The anomaly, in this formulation, is the residue at the pole.

\section{Decomposition of the correlator}\label{sectiondecompcorr}
In this section we find the most general expression of the $\langle TTJ_5\rangle$ correlator, satisfying the anomalous conservation WI and trace WI.
The analysis is performed by applying the L/T decomposition to the correlator.
We focus on a parity odd four-dimensional correlator, therefore its tensorial structure will involve the antisymmetric tensor $\varepsilon^{\mu\nu\rho\sigma}$.\\
We start by decomposing the energy-momentum tensor $T^{\m \n}$ and the current $J_5^\m$ in terms of their transverse-traceless part and longitudinal ones (also called "local")
\begin{align}
	T^{\mu_i\nu_i}(p_i)&= t^{\mu_i\nu_i}(p_i)+t_{loc}^{\mu_i\nu_i}(p_i),\label{decT}\\
	J_5^{\mu_i}(p_i)&= j_5^{\mu_i}(p_i)+j_{5 \, loc}^{\mu_i}(p_i),\label{decJ}
\end{align}
where
\begin{align}
	\label{loct}
	&t^{\mu_i\nu_i}(p_i)=\Pi^{\mu_i\nu_i}_{\alpha_i\beta_i}(p_i)\,T^{\alpha_i \beta_i}(p_i), &&t_{loc}^{\mu_i\nu_i}(p_i)=\Sigma^{\mu_i\nu_i}_{\alpha_i\beta_i}(p)\,T^{\alpha_i \beta_i}(p_i),\nn\\
	&j_5^{\mu_i}(p_i)=\pi^{\mu_i}_{\alpha_i}(p_i)\,J_5^{\alpha_i }(p_i), \hspace{1ex}&&j_{5\, loc}^{\mu_i}(p_i)=\frac{p_i^{\mu_i}\,p_{i\,\alpha_i}}{p_i^2}\,J_5^{\alpha_i}(p_i),
\end{align}
having introduced the transverse-traceless ($\Pi$), transverse $(\pi)$ and longitudinal ($\Sigma$) projectors, given respectively by 
\begin{align}
	\label{prozero}
	&\pi^{\mu}_{\alpha}  = \delta^{\mu}_{\alpha} - \frac{p^{\mu} p_{\alpha}}{p^2}, \\&
	\Pi^{\mu \nu}_{\alpha \beta}  = \frac{1}{2} \left( \pi^{\mu}_{\alpha} \pi^{\nu}_{\beta} + \pi^{\mu}_{\beta} \pi^{\nu}_{\alpha} \right) - \frac{1}{d - 1} \pi^{\mu \nu}\pi_{\alpha \beta}\label{TTproj}, \\&
	\Sigma^{\mu_i\nu_i}_{\alpha_i\beta_i}=\frac{p_{i\,\beta_i}}{p_i^2}\Big[2\delta^{(\nu_i}_{\alpha_i}p_i^{\mu_i)}-\frac{p_{i\alpha_i}}{(d-1)}\left(\delta^{\mu_i\nu_i}+(d-2)\frac{p_i^{\mu_i}p_i^{\nu_i}}{p_i^2}\right)\Big]+\frac{\pi^{\mu_i\nu_i}(p_i)}{(d-1)}\delta_{\alpha_i\beta_i}\label{Lproj}.
\end{align}
Such decomposition allows to split our correlation function into the following terms
\begin{equation} \label{eq:splitlongttpart}
	\begin{aligned}
		\left\langle T^{\mu_{1} \n_{1}} T^{\mu_{2} \n_2} J_5^{\mu_{3}}\right\rangle=&\left\langle t^{\mu_{1} \n_{1}} t^{\mu_{2}\n_2} j_5^{\mu_{3}}\right\rangle+\left\langle T^{\mu_{1} \n_{1}} T^{\mu_{2}\n_2} j_{5\, l o c}^{\mu_{3}}\right\rangle+\left\langle T^{\mu_{1} \n_{1}} t_{l o c}^{\mu_{2}\n_2} J_5^{\mu_{3}}\right\rangle+\left\langle t_{l o c}^{\mu_{1} \n_{1}} T^{\mu_{2}\n_2} J_5^{\mu_{3}}\right\rangle \\
		&-\left\langle T^{\mu_{1} \n_{1}} t_{l o c}^{\mu_{2}\n_2} j_{5\, l o c}^{\mu_{3}}\right\rangle-\left\langle t_{l o c}^{\mu_{1} \n_{1}} t_{l o c}^{\mu_{2}\n_2} J_5^{\mu_{3}}\right\rangle-\left\langle t_{l o c}^{\mu_{1} \n_{1}} T^{\mu_{2}\n_2} j_{5\, l o c}^{\mu_{3}}\right\rangle+\left\langle t_{l o c}^{\mu_{1} \n_{1}} t_{l o c}^{\mu_{2}\n_2} j_{5\, l o c}^{\mu_{3}}\right\rangle .
	\end{aligned}
\end{equation}
Using the conservation and trace WIs derived in the previous section, it is then possible to completely fix all the longitudinal parts, i.e. the terms containing at least one $t^{\m \n}_{{loc}}$ or $j^\m_{5\, {loc}}$. We start by considering the non-anomalous equations
\begin{equation} \label{ppp}
	\begin{aligned}
		&\delta_{\mu_i\nu_i }\braket{T^{\mu_1\nu_1}(p_1)T^{\mu_2\nu_2}(p_2)J_5^{\mu_3}(p_3)}=0,\qquad \quad&& i=\{1,2\}\\
		&p_{i\mu_i}\,\braket{T^{\mu_1\nu_1}(p_1)T^{\mu_2\nu_2}(p_2)J_5^{\mu_3}(p_3)}=0,\qquad \quad && i=\{1,2\}.
	\end{aligned}
\end{equation}
Thanks to these WIs, we can eliminate most of terms on the right-hand side of equation \eqref{eq:splitlongttpart}, ending up only with two terms
\begin{equation}
	\left\langle T^{\mu_{1} \n_{1}} T^{\mu_{2} \n_2} J_5^{\mu_{3}}\right\rangle=\left\langle t^{\mu_{1} \n_{1}} t^{\mu_{2}\n_2} j_5^{\mu_{3}}\right\rangle+\left\langle T^{\mu_{1} \n_{1}} T^{\mu_{2}\n_2} j_{5\, l o c}^{\mu_{3}}\right\rangle=\left\langle t^{\mu_{1} \n_{1}} t^{\mu_{2}\n_2} j_5^{\mu_{3}}\right\rangle+\left\langle t^{\mu_{1} \n_{1}} t^{\mu_{2}\n_2} j_{5\, l o c}^{\mu_{3}}\right\rangle.
\end{equation}
The remaining local term is then fixed by the anomalous WI of $J_5$.
First, we construct the most general expression in terms of tensorial structures and form factors
\beq \label{eq:locpartdecompgen}
\left\langle t^{\mu_{1} \n_{1}} t^{\mu_{2}\n_2} j_{5\, l o c}^{\mu_{3}}\right\rangle=
p_3^{\m_3}\, \Pi^{\m_1 \n_1}_{\a_1  \b_1}(p_1) \, \Pi^{\m_2 \n_2}_{\a_2 \b_2}(p_2) \,  \varepsilon^{\a_1 \a_2 p_1 p_2}\left( F_1 \, g^{\b_1 \b_2} + F_2 \, p_1^{\b_2} p_2^{\b_1} \right)
\eeq
where, due to the Bose symmetry, both $F_1$ and $F_2$ are symmetric under the exchange $\left(p_1\leftrightarrow p_2\right)$. Then, recalling the definition of $j_{5\, l o c}$ and the anomalous WI 
\small
\begin{equation}\label{eq:idwanomlp3}
	p_{3\mu_3}\braket{T^{\mu_1\nu_1}(p_1)T^{\mu_2\nu_2}(p_2)J_5^{\mu_3}(p_3)}= 4\, i \, a_2 \, (p_1 \cdot p_2) \left\{ \left[\varepsilon^{\nu_1 \nu_2 p_1 p_2}\left(g^{\mu_1 \mu_2}- \frac{p_1^{\mu_2} p_2^{\mu_1}}{p_1 \cdot p_2}\right) +\left( \mu_1 \leftrightarrow \nu_1 \right) \right] +\left( \mu_2 \leftrightarrow \nu_2 \right) \right\},
\end{equation}
\normalsize
we can write
\begin{equation} \label{eq:anompolettj}
	\left\langle t^{\mu_{1} \n_{1}} t^{\mu_{2}\n_2} j_{5\, l o c}^{\mu_{3}}\right\rangle= 4 ia_2 \frac{p_3^{\mu_3}}{p_3^2} \, (p_1 \cdot p_2) \left\{ \left[\varepsilon^{\nu_1 \nu_2 p_1 p_2}\left(g^{\mu_1 \mu_2}- \frac{p_1^{\mu_2} p_2^{\mu_1}}{p_1 \cdot p_2}\right) +\left( \mu_1 \leftrightarrow \nu_1 \right) \right] +\left( \mu_2 \leftrightarrow \nu_2 \right) \right\}.
\end{equation}
One can show that this formula coincides with eq$.$ \eqref{eq:locpartdecompgen} after contracting the projectors' indices and fixing the form factors in the following way
\begin{equation}
	\begin{aligned}
		&F_1=\frac{16 i a_2 (p_1\cdot p_2)}{p_3^2}, \qquad\qquad\qquad
		F_2=-\frac{16 i a_2}{p_3^2}.
	\end{aligned}
\end{equation}
Therefore, all the local terms of the $\langle TTJ_5\rangle$ are fixed. The only remaining term to be studied in order to reconstruct the entire correlator is the transverse-traceless part $\left\langle t^{\mu_{1} \n_{1}} t^{\mu_{2}\n_2} j_5^{\mu_{3}}\right\rangle$.
Its explicit form is given by
\begin{equation} \label{eq:defX}
	\left\langle t^{\mu_1 \n_1}\left(p_1\right) t^{\mu_2 \n_2}\left(p_2\right) j_5^{\mu_3}\left(p_3\right)\right\rangle=\Pi_{\alpha_1 \beta_1}^{\mu_1 \nu_1}\left(p_1\right) \Pi_{\alpha_2 \b_2}^{\mu_2 \n_2}\left(p_2\right) \pi_{\alpha_3}^{\mu_3}\left(p_3\right) X^{\alpha_1 \beta_1 \alpha_2 \b_2 \alpha_3}
\end{equation}
where $X^{\alpha_1 \beta_1 \alpha_2 \b_2 \alpha_3}$ is a general rank five tensor built by products of metric tensors, momenta and the Levi-Civita symbol
with the appropriate choice of indices. Indeed, as a consequence of the projectors in \eqref{eq:defX}, $X^{\alpha_1 \beta_1 \alpha_2 \b_2 \alpha_3}$ can
not be constructed by using $g_{\a_i \b_i}$, nor by $p_{i \, \a_i}$ with $i =\{1,2,3\}$. 
We also must keep in mind that, due to symmetries of the correlator, form factors associated with structures linked by a $(1\leftrightarrow 2)$ transformation (the gravitons exchange) are dependent. Then, the transverse-traceless part can be written as
\begin{equation}\label{eq:decttnm}
	\begin{aligned}
		\langle t^{\mu_{1} \nu_{1}}\left({p}_{1}\right)& t^{\mu_{2} \nu_{2}}\left({p}_{2}\right) j_5^{\mu_{3} }\left({p_3}\right)\rangle=\Pi_{\alpha_{1} \beta_{1}}^{\mu_{1} \nu_{1}}\left({p}_{1}\right) \Pi_{\alpha_{2} \beta_{2}}^{\mu_{2} \nu_{2}}\left({p}_{2}\right) \pi_{\alpha_{3}}^{\mu_{3}}\left({p_3}\right) \bigg[
		\\
		&A_1\varepsilon^{p_1\alpha_1\alpha_2\alpha_3}p_2^{\beta_1}p_3^{\beta_2}
		-A_1(p_1\leftrightarrow p_2) \varepsilon^{p_2\alpha_1\alpha_2\alpha_3}p_2^{\beta_1}p_3^{\beta_2}\\
		&+A_2\varepsilon^{p_1\alpha_1\alpha_2\alpha_3}\delta^{\beta_1\beta_2}-
		A_2(p_1\leftrightarrow p_2)\varepsilon^{p_2\alpha_1\alpha_2\alpha_3}\delta^{\beta_1\beta_2}\\
		&+A_3\varepsilon^{p_1p_2\alpha_1\alpha_2}p_2^{\beta_1}p_3^{\beta_2}p_1^{\alpha_3}
		+A_4\varepsilon^{p_1p_2\alpha_1\alpha_2}\delta^{\beta_1\beta_2}p_1^{\alpha_3}\\
		&+A_5 \varepsilon^{p_1 p_2\alpha_1 \alpha_3 } p_2^{\beta_1}p_3^{\alpha_2} p_3^{\b_2} +A_5(p_1\leftrightarrow p_2) \varepsilon^{p_1 p_2 \a_2 \a_3 } p_3^{\beta_2}p_2^{\alpha_1} p_2^{\beta_1}  \\
		&+A_6\varepsilon^{p_1 p_2 \alpha_1 \alpha_3} p_3^{\alpha_2}\d^{\beta_1\beta_2} +A_6(p_1\leftrightarrow p_2)\varepsilon^{p_1 p_2 \alpha_2 \alpha_3 }p_2^{\a_1}\delta^{\b_1\beta_2} \\
		&+A_7\varepsilon^{p_1 p_2\alpha_1 \alpha_2 } p_2^{\beta_1} \delta^{\beta_2 \alpha_3}
		-A_7(p_1\leftrightarrow p_2) \varepsilon^{p_1 p_2\alpha_1 \alpha_2 } p_3^{\b_2} \d^{\b_1 \a_3} \bigg]
	\end{aligned}
\end{equation}
where $A_3$ and $A_4$ are antisymmetric under the exchange $(p_1\leftrightarrow p_2)$
and we have made a choice on which independent momenta to consider for each index
\begin{equation}
	\{ \alpha_1, \beta_1\} \leftrightarrow p_2, 
	\qquad\{ \alpha_2 ,\beta_2\} \leftrightarrow p_3, 
	\qquad\{ \alpha_3\} \leftrightarrow p_1.
\end{equation}
Since we are working in $d = 4$ the form factors in eq$.$ \eqref{eq:decttnm} are not all independent and the decomposition is not minimal. Indeed, one needs to consider the following class of tensor identities
\begin{equation}\label{eq:classScID}
	0=\varepsilon^{[p_1p_2\alpha_1\alpha_2}\delta^{\alpha_3]}_\alpha
\end{equation}
If we set
$\alpha=\beta_1$ or $\alpha=\beta_2$ and apply the projectors, we have
\begin{equation}
	\begin{aligned}
		\Pi_{\alpha_{1} \beta_{1}}^{\mu_{1} \nu_{1}}\left({p}_{1}\right) \Pi_{\alpha_{2} \beta_{2}}^{\mu_{2} \nu_{2}}\left({p}_{2}\right) &\pi_{\alpha_{3}}^{\mu_{3}}\left({p_3}\right) \bigg[\varepsilon^{p_1 p_2\alpha_1\alpha_2} \delta^{\alpha_3\beta_1}\bigg]=\\&
		\Pi_{\alpha_{1} \beta_{1}}^{\mu_{1} \nu_{1}}\left({p}_{1}\right) \Pi_{\alpha_{2} \beta_{2}}^{\mu_{2} \nu_{2}}\left({p}_{2}\right) \pi_{\alpha_{3}}^{\mu_{3}}\left({p_3}\right) \bigg[\varepsilon^{p_1\alpha_1\alpha_2\alpha_3} p_2^{\beta_1}+\varepsilon^{p_1 p_2\alpha_1\alpha_3} \delta^{\alpha_2 \beta_1}\bigg],\\ \Pi_{\alpha_{1} \beta_{1}}^{\mu_{1} \nu_{1}}\left({p}_{1}\right) \Pi_{\alpha_{2} \beta_{2}}^{\mu_{2} \nu_{2}}\left({p}_{2}\right) &\pi_{\alpha_{3}}^{\mu_{3}}\left({p_3}\right)\bigg[
		\varepsilon^{p_1 p_2\alpha_1\alpha_2} \delta^{\alpha_3\beta_2}\bigg]= \\ &
		\Pi_{\alpha_{1} \beta_{1}}^{\mu_{1} \nu_{1}}\left({p}_{1}\right) \Pi_{\alpha_{2} \beta_{2}}^{\mu_{2} \nu_{2}}\left({p}_{2}\right) \pi_{\alpha_{3}}^{\mu_{3}}\left({p_3}\right) \bigg[\varepsilon^{p_2\alpha_1\alpha_2\alpha_3} p_3^{\beta_2}-\varepsilon^{p_1 p_2\alpha_2\alpha_3} \delta^{\alpha_1\beta_2}\bigg]
	\end{aligned}
\end{equation}
according to which we can rewrite the tensorial structures multiplying $A_7$ in terms of the others.
If we instead contract the identity \eqref{eq:classScID} with $p_{1\alpha}$ and $p_{2\alpha}$, we arrive to
\begin{equation}
	\begin{aligned}
		\Pi_{\alpha_{1} \beta_{1}}^{\mu_{1} \nu_{1}}\left({p}_{1}\right) \Pi_{\alpha_{2} \beta_{2}}^{\mu_{2} \nu_{2}}\left({p}_{2}\right) &\pi_{\alpha_{3}}^{\mu_{3}}\left({p_3}\right) \bigg[ \varepsilon^{p_1 p_2\alpha_1\alpha_3} p_3^{\alpha_2}\bigg]=\\&
		\Pi_{\alpha_{1} \beta_{1}}^{\mu_{1} \nu_{1}}\left({p}_{1}\right) \Pi_{\alpha_{2} \beta_{2}}^{\mu_{2} \nu_{2}}\left({p}_{2}\right) \pi_{\alpha_{3}}^{\mu_{3}}\left({p_3}\right)
		\bigg[ \varepsilon^{p_1\alpha_1\alpha_2\alpha_3} \left(p_1\cdot p_2 \right)-\varepsilon^{p_1 p_2\alpha_1\alpha_2} p_1^{\alpha_3}-\varepsilon^{p_2\alpha_1\alpha_2
			\alpha_3} p_1^2\bigg],\\
		\Pi_{\alpha_{1} \beta_{1}}^{\mu_{1} \nu_{1}}\left({p}_{1}\right) \Pi_{\alpha_{2} \beta_{2}}^{\mu_{2} \nu_{2}}\left({p}_{2}\right)& \pi_{\alpha_{3}}^{\mu_{3}}\left({p_3}\right) \bigg[
		\varepsilon^{p_1 p_2\alpha_2\alpha_3} p_2^{\alpha_1}\bigg]=\\&
		\Pi_{\alpha_{1} \beta_{1}}^{\mu_{1} \nu_{1}}\left({p}_{1}\right) \Pi_{\alpha_{2} \beta_{2}}^{\mu_{2} \nu_{2}}\left({p}_{2}\right) \pi_{\alpha_{3}}^{\mu_{3}}\left({p_3}\right) \bigg[
		\varepsilon^{p_1\alpha_1\alpha_2\alpha_3} p_2^2+\varepsilon^{p_1 p_2\alpha_1\alpha_2} p_1^{\alpha_3}- \varepsilon^{p_2\alpha_1\alpha_2\alpha_3} \left(p_1\cdot p_2\right)\bigg]
	\end{aligned}	
\end{equation}
according to which we can rewrite the form factors $A_5$ and $A_6$ in terms of the first four.
We conclude that the general structure of the transverse-traceless part is given by
\begin{equation}\label{eq:decomptt}
	\begin{aligned}
		\langle t^{\mu_{1} \nu_{1}}\left({p}_{1}\right)& t^{\mu_{2} \nu_{2}}\left({p}_{2}\right) j_5^{\mu_{3} }\left({p_3}\right)\rangle=\Pi_{\alpha_{1} \beta_{1}}^{\mu_{1} \nu_{1}}\left({p}_{1}\right) \Pi_{\alpha_{2} \beta_{2}}^{\mu_{2} \nu_{2}}\left({p}_{2}\right) \pi_{\alpha_{3}}^{\mu_{3}}\left({p_3}\right) \bigg[
		\\
		&A_1\varepsilon^{p_1\alpha_1\alpha_2\alpha_3}p_2^{\beta_1}p_3^{\beta_2}
		-A_1(p_1\leftrightarrow p_2) \varepsilon^{p_2\alpha_1\alpha_2\alpha_3}p_2^{\beta_1}p_3^{\beta_2}\\
		&+A_2\varepsilon^{p_1\alpha_1\alpha_2\alpha_3}\delta^{\beta_1\beta_2}-
		A_2(p_1\leftrightarrow p_2)\varepsilon^{p_2\alpha_1\alpha_2\alpha_3}\delta^{\beta_1\beta_2}\\
		&+A_3\varepsilon^{p_1p_2\alpha_1\alpha_2}p_2^{\beta_1}p_3^{\beta_2}p_1^{\alpha_3}
		+A_4\varepsilon^{p_1p_2\alpha_1\alpha_2}\delta^{\beta_1\beta_2}p_1^{\alpha_3}
		\bigg]
	\end{aligned}
\end{equation}
where we have redefined the form factors $A_1, \dots, A_4$. Once again, $A_3$ and $A_4$ are antisymmetric under the exchange $(p_1\leftrightarrow p_2)$.

\section{The conformal analysis of the $\langle TTJ_5 \rangle $}
In the previous section we have seen that the conservation and trace WIs fix the longitudinal part of the correlator. In this section we examine the conformal constraints on the $\langle TTJ_5 \rangle $, following closely the methodology adopted in \cite{Bzowski:2013sza}. We will see that the transverse-traceless part of the correlator is completely determined by conformal invariance together with the $R\tilde{R}$ part of the boundary condition coming from the anomaly relation \eqref{eq:anomaliachirale}, corresponding to the anomalous coefficient $a_2$.

\subsection{Dilatation Ward identities}
The invariance of the correlator under dilatation is reflected in the equation \eqref{eqD}.
 Due to this constraint, the transverse-traceless part of the correlator has to satisfy the equation
\begin{align}\label{eq:Diltt}
	\left(\sum_{i=1}^3\Delta_i-2d-\sum_{i=1}^2\,p_i^\mu\frac{\partial}{\partial p_i^\mu}\right)\braket{t^{\mu_1\nu_1}(p_1)t^{\mu_2\nu_2}(p_2)j_5^{\mu_3} (p_3)}=0.
\end{align}
By using the chain rule
\begin{align}
	\frac{\partial}{\partial p_i^\mu}=\sum_{j=1}^3\frac{\partial p_j}{\partial p_i^\mu}\frac{\partial}{\partial p_j}
\end{align}
to express the derivatives respect to 4-vectors in term of the invariants $p_i=|\sqrt{p_i^2}|$, we rewrite \eqref{eq:Diltt} as a constraint on the form factors 
\begin{align}
	\sum_{i=1}^{3} p_i \frac{\partial A_j}{\partial p_i }-\left(\sum_{i=1}^3\Delta_i-2d- N_j\right) A_j=0
\end{align}
with $N_j$ the number of momenta that the form factors multiply in the decomposition of eq.  \eqref{eq:decomptt} 
\begin{align}
	N_1=3,\qquad\qquad N_2=1,\qquad\qquad N_3=5, \qquad\qquad N_4=3,
\end{align}

\subsection{Special conformal Ward identities}

The invariance of the correlator with respect to the special conformal transformations is encoded in the following equation
\begin{equation}
	\begin{aligned}
		0=&\, \mathcal{K}^\kappa\left\langle T^{\mu_1 \nu_1}\left(p_1\right) T^{\mu_2 \nu_2}\left(p_2\right) J_5^{\mu_3 }\left(p_3\right)\right\rangle \\
		&\equiv 
		\sum_{j=1}^2\left(2\left(\Delta_j-d\right) \frac{\partial}{\partial p_{j \kappa}}-2 p_j^\alpha \frac{\partial}{\partial p_j^\alpha} \frac{\partial}{\partial p_{j \kappa}}+\left(p_j\right)^\kappa \frac{\partial}{\partial p_j^\alpha} \frac{\partial}{\partial p_{j \alpha}}\right)\left\langle T^{\mu_1 \nu_1}\left(p_1\right) T^{\mu_2 \nu_2}\left(p_2\right) J_5^{\mu_3}\left(p_3\right)\right\rangle \\
		&\qquad+4\left(\delta^{\kappa\left(\mu_1\right.} \frac{\partial}{\partial p_1^{\alpha_1}}-\delta_{\alpha_1}^\kappa \delta_\lambda^{\left(\mu_1\right.} \frac{\partial}{\partial p_{1 \lambda}}\right)\left\langle T^{\left.\nu_1\right) \alpha_1}\left(p_1\right) T^{\mu_2 \nu_2}\left(p_2\right) J_5^{\mu_3 }\left(p_3\right)\right\rangle \\
		&\qquad+4\left(\delta^{\kappa\left(\mu_2\right.} \frac{\partial}{\partial p_2^{\alpha_2}}-\delta_{\alpha_2}^\kappa \delta_\lambda^{\left(\mu_2\right.} \frac{\partial}{\partial p_{2 \lambda}}\right)\left\langle T^{\left.\nu_2\right) \alpha_2}\left(p_2\right) T^{\mu_1 \nu_1}\left(p_1\right) J_5^{\mu_3 }\left(p_3\right)\right\rangle .
	\end{aligned}
\end{equation}
The special conformal operator $\mathcal{K}^\kappa$ acts as an endomorphism on the transverse-traceless sector of the entire correlator. 
Therefore we can perform a transverse-traceless projection on all the indices in order to identify a set of partial differential equations 
\begin{align}
	0=\Pi^{\rho_{1} \sigma_{1}}_{\mu_{1} \nu_{1}}&\left({p}_{1}\right) \Pi^{\rho_{2} \sigma_{2}}_{\mu_{2} \nu_{2}}\left({p}_{2}\right) \pi^{\rho_{3}}_{\mu_{3}}\left({p_3}\right) 
	\mathcal{K}^\kappa 
	\left\langle T^{\mu_1 \nu_1}\left(p_1\right) T^{\mu_2 \nu_2}\left(p_2\right) J_5^{\mu_3}\left(p_3\right)\right\rangle
	=\nn \\&
	\Pi^{\rho_{1} \sigma_{1}}_{\mu_{1} \nu_{1}}\left({p}_{1}\right) \Pi^{\rho_{2} \sigma_{2}}_{\mu_{2} \nu_{2}}\left({p}_{2}\right) \pi^{\rho_{3}}_{\mu_{3}}\left({p_3}\right)  \mathcal{K}^\kappa 
	\bigg(
	\left\langle t^{\mu_{1} \n_{1}} t^{\mu_{2}\n_2} j_5^{\mu_{3}}\right\rangle+\left\langle t^{\mu_{1} \n_{1}} t^{\mu_{2}\n_2} j_{5\, l o c}^{\mu_{3}}\right\rangle\bigg),
\end{align}
splitting the correlator into its transverse and longitudinal parts.  
The action of the special conformal operator $\mathcal{K}^\kappa$ on the longitudinal part of the correlator is given by
\begin{equation}
	\begin{aligned}
		&\Pi^{\rho_{1} \sigma_{1}}_{\mu_{1} \nu_{1}}\left({p}_{1}\right) \Pi^{\rho_{2} \sigma_{2}}_{\mu_{2} \nu_{2}}\left({p}_{2}\right) \pi^{\rho_{3}}_{\mu_{3}}\left({p_3}\right) \bigg[ \mathcal{K}^\kappa 
		\left\langle t^{\mu_{1} \n_{1}} t^{\mu_{2}\n_2} j_{5\, l o c}^{\mu_{3}}\right\rangle\bigg]=\\&\qquad\qquad
		\Pi^{\rho_{1} \sigma_{1}}_{\mu_{1} \nu_{1}}\left({p}_{1}\right) \Pi^{\rho_{2} \sigma_{2}}_{\mu_{2} \nu_{2}}\left({p}_{2}\right) \pi^{\rho_{3}}_{\mu_{3}}\left({p_3}\right) \bigg[ 2\frac{(\Delta_3-1)}{p_3^2}\delta^{\kappa \mu_3}p_{3\alpha}	\left\langle t^{\mu_{1} \n_{1}} t^{\mu_{2}\n_2} j_{5\, l o c}^{\alpha}\right\rangle \bigg].
	\end{aligned}
\end{equation}
Using the eq. \eqref{eq:idwanomlp3} together with the Schouten identities mentioned in the Appendix \ref{appendix:Schouten}, we can write
\small
\begin{equation} \label{eq:anomcontrscwi}
	\begin{aligned}		
		&	\Pi^{\rho_{1} \sigma_{1}}_{\mu_{1} \nu_{1}}\left({p}_{1}\right) \Pi^{\rho_{2} \sigma_{2}}_{\mu_{2} \nu_{2}}\left({p}_{2}\right) \pi^{\rho_{3}}_{\mu_{3}}\left({p_3}\right)\bigg[ \mathcal{K}^\kappa 
		\left\langle t^{\mu_{1} \n_{1}} t^{\mu_{2}\n_2} j_{5\, l o c}^{\mu_{3}}\right\rangle\bigg]
		=
		\Pi^{\rho_{1} \sigma_{1}}_{\mu_{1} \nu_{1}}\left({p}_{1}\right) \Pi^{\rho_{2} \sigma_{2}}_{\mu_{2} \nu_{2}}\left({p}_{2}\right) \pi^{\rho_{3}}_{\mu_{3}}\left({p_3}\right)
		\frac{16 \,i\, a_2 \,(\Delta_3-1)}{p_3^2}
		\bigg[ \\&
		p_1^\kappa \varepsilon^{p_2\mu_1\mu_2\mu_3} \bigg(-2 p_2^{\nu_1}p_3^{\nu_2}+
		(p_1^2+p_2^2-p_3^2) \delta^{\nu_1\nu_2}\bigg)
		+ p_2^\kappa \varepsilon^{p_1\mu_1\mu_2\mu_3} \bigg( 2 p_2^{\nu_1}p_3^{\nu_2}
		- (p_1^2+p_2^2-p_3^2)\delta^{\nu_1\nu_2}
		\bigg)\\&
		\delta^{\kappa\mu_1}\bigg(
		+(p_1^2+p_2^2-p_3^2)\varepsilon^{p_1p_2\mu_2\mu_3}\delta^{\nu_1\nu_2}
		-2p_2^2\varepsilon^{p_1\mu_2\mu_3\nu_1}p_3^{\nu_2}
		-(p_1^2+p_2^2-p_3^2)\varepsilon^{p_2\mu_2\mu_3\nu_1}p_3^{\nu_2}
		+2\varepsilon^{p_1p_2\mu_2\nu_1}p_1^{\mu_3}p_3^{\nu_2}\bigg)
		\\&\delta^{\kappa\mu_2}\bigg(
		-(p_1^2+p_2^2-p_3^2)\varepsilon^{p_1p_2\mu_1\mu_3}\delta^{\nu_1\nu_2}
		+2p_1^2\varepsilon^{p_2\mu_1\mu_3\nu_2}p_2^{\nu_1}
		+(p_1^2+p_2^2-p_3^2)\varepsilon^{p_1\mu_1\mu_3\nu_2}p_2^{\nu_1}
		-2\varepsilon^{p_1p_2\mu_1\nu_2}p_1^{\mu_3}p_2^{\nu_1}\bigg)
		\bigg].
	\end{aligned}
\end{equation}
\normalsize
Using the Schouten identities reported in Appendix \ref{appendix:Schouten}, we can then decompose the action of the special conformal operator on the entire correlator in the following minimal expression
\small
\begin{equation}
	\begin{aligned}
		0=	\Pi^{\rho_{1} \sigma_{1}}_{\mu_{1} \nu_{1}}&\left({p}_{1}\right) \Pi^{\rho_{2} \sigma_{2}}_{\mu_{2} \nu_{2}}\left({p}_{2}\right) \pi^{\rho_{3}}_{\mu_{3}}\left({p_3}\right)
		\bigg(\mathcal{K}^\kappa 
		\left\langle T^{\mu_1 \nu_1}\left(p_1\right) T^{\mu_2 \nu_2}\left(p_2\right) J_5^{\mu_3}\left(p_3\right)\right\rangle
		\bigg)=
		\Pi^{\rho_{1} \sigma_{1}}_{\mu_{1} \nu_{1}}\left({p}_{1}\right) \Pi^{\rho_{2} \sigma_{2}}_{\mu_{2} \nu_{2}}\left({p}_{2}\right) \pi^{\rho_{3}}_{\mu_{3}}\left({p_3}\right) \bigg[\\&
		p_1^\kappa \bigg(
		C_{11}\varepsilon^{p_1\mu_1\mu_2\mu_3}p_2^{\nu_1}p_3^{\nu_2}+
		C_{12}\varepsilon^{p_2\mu_1\mu_2\mu_3}p_2^{\nu_1}p_3^{\nu_2}+
		C_{13}\varepsilon^{p_1\mu_1\mu_2\mu_3}\delta^{\nu_1\nu_2}
		+C_{14}\varepsilon^{p_2\mu_1\mu_2\mu_3}\delta^{\nu_1\nu_2}\\&
		+C_{15}\varepsilon^{p_1p_2\mu_1\mu_2}p_2^{\nu_1}p_3^{\nu_2}p_1^{\mu_3}
		+C_{16}\varepsilon^{p_1p_2\mu_1\mu_2}\delta^{\nu_1\nu_2}p_1^{\mu_3}\bigg)\\&
		+p_2^\kappa \bigg(
		C_{21}\varepsilon^{p_1\mu_1\mu_2\mu_3}p_2^{\nu_1}p_3^{\nu_2}+
		C_{22}\varepsilon^{p_2\mu_1\mu_2\mu_3}p_2^{\nu_1}p_3^{\nu_2}+
		C_{23}\varepsilon^{p_1\mu_1\mu_2\mu_3}\delta^{\nu_1\nu_2}
		+C_{24}\varepsilon^{p_2\mu_1\mu_2\mu_3}\delta^{\nu_1\nu_2}\\&
		+C_{25}\varepsilon^{p_1p_2\mu_1\mu_2}p_2^{\nu_1}p_3^{\nu_2}p_1^{\mu_3}
		+C_{26}\varepsilon^{p_1p_2\mu_1\mu_2}\delta^{\nu_1\nu_2}p_1^{\mu_3}\bigg)\\&
		+\delta^{\kappa\mu_1}\bigg(
		C_{31}\varepsilon^{p_1\mu_2\mu_3\nu_1}p_3^{\nu_2}
		+C_{32}\varepsilon^{p_2\mu_2\mu_3\nu_1}p_3^{\nu_2}
		+C_{33}\varepsilon^{p_1p_2\mu_2\nu_1}p_1^{\mu_3}p_3^{\nu_2}
		+C_{34}\varepsilon^{p_1p_2\mu_2\mu_3}\delta^{\nu_1\nu_2}
		\bigg)
		\\&
		+\delta^{\kappa\mu_2}\bigg(
		C_{41}\varepsilon^{p_1\mu_1\mu_3\nu_2}p_2^{\nu_1}
		+C_{42}\varepsilon^{p_2\mu_1\mu_3\nu_2}p_2^{\nu_1}
		+C_{43}\varepsilon^{p_1p_2\mu_1\nu_2}p_1^{\mu_3}p_2^{\nu_1}
		+C_{44}\varepsilon^{p_1p_2\mu_1\mu_3}\delta^{\nu_1\nu_2}
		\bigg)
		\\&
		+C_{51}\varepsilon^{\kappa\mu_1\mu_2\mu_3}\delta^{\nu_1\nu_2}
		+C_{52}\varepsilon^{\kappa\mu_1\mu_2\mu_3}p_2^{\nu_1}p_3^{\nu_2}
		+C_{53}\varepsilon^{p_1\kappa\mu_1\mu_2}p_1^{\mu_3}\delta^{\nu_1\nu_2}
		+C_{54}\varepsilon^{p_2\kappa\mu_1\mu_2}p_1^{\mu_3}\delta^{\nu_1\nu_2}\bigg]
	\end{aligned}
\end{equation}
\normalsize
where the coefficients $C_{ij}$ depend on the gravitational anomalous coefficient $a_2$, the form factors $A_i$ and their derivatives with respect to the momenta.\\
Due to the independence of the tensorial structures listed in the equation above, all the coefficients $C_{ij}$ need to vanish. In particular the primary equations are
\begin{equation} \label{eq:primeqscwi}
	\begin{aligned}
		0=C_{ij} \qquad\qquad\qquad\qquad i=\{1,2\} ,\quad j=\{1,\dots 6\}\\
	\end{aligned}
\end{equation}
They correspond to second order differential equations.
The secondary equations are instead given by
\begin{equation}
	\begin{aligned}
		0=C_{ij} \qquad\qquad\qquad\qquad i=\{3,4,5\}, \quad j=\{1,\dots 4\}\\
	\end{aligned}
\end{equation}
and they are differential equations of the first order.

\subsection{Solving the CWIs}

The most general solution of the CWIs of the $\langle TTJ_5\rangle$ can be written in terms of integrals involving a product of three Bessel functions, namely 3K integrals \cite{Bzowski:2013sza}. For a detailed review on the properties of such integrals, see also \cite{Bzowski:2015yxv}.
We recall the definition of the general 3K integral
\begin{equation}
	I_{\alpha\left\{\beta_1, \beta_2, \beta_3\right\}}\left(p_1, p_2, p_3\right)=\int d x x^\alpha \prod_{j=1}^3 p_j^{\beta_j} K_{\beta_j}\left(p_j x\right)
\end{equation}
where $K_\nu$ is a modified Bessel function of the second kind 
\begin{equation}
	K_\nu(x)=\frac{\pi}{2} \frac{I_{-\nu}(x)-I_\nu(x)}{\sin (\nu \pi)}, \qquad \nu \notin \mathbb{Z} \qquad\qquad I_\nu(x)=\left(\frac{x}{2}\right)^\nu \sum_{k=0}^{\infty} \frac{1}{\Gamma(k+1) \Gamma(\nu+1+k)}\left(\frac{x}{2}\right)^{2 k}
\end{equation}
with the property
\begin{equation}
	K_n(x)=\lim _{\epsilon \rightarrow 0} K_{n+\epsilon}(x), \quad n \in \mathbb{Z}
\end{equation}
We will also use the reduced version of the 3K integral defined as
\begin{equation}
	J_{N\left\{k_j\right\}}=I_{\frac{d}{2}-1+N\left\{\Delta_j-\frac{d}{2}+k_j\right\}}
\end{equation}
where we introduced the condensed notation $\{k_j \} = \{k_1, k_2, k_3 \}$.
The 3K integrals satisfy an equation analogous to the dilatation equation with scaling degree \cite{Bzowski:2013sza}
\begin{equation}
	\text{deg}\left(J_{N\left\{k_j\right\}}\right)=\Delta_t+k_t-2 d-N
\end{equation}
where 
\begin{equation}
	k_t=k_1+k_2+k_3,\qquad\qquad \Delta_t=\Delta_1+\Delta_2+\Delta_3
\end{equation}
From this analysis, it is simple to relate the form factors to the 3K integrals. Indeed, the dilatation WI tells us that the form factors $A_i$ can to be written as a combination of integrals of the following type
\begin{equation}
	J_{N_i+k_t,\{k_1,k_2,k_3\}}
\end{equation}
where $N_i$ is the number of momenta that the form factors multiplies in the decomposition \eqref{eq:decomptt}. The special CWIs fix the remaining indices $k_1$, $k_2$ and $k_3$.\\
We start by considering the explicit form of the primary equations \eqref{eq:primeqscwi} involving the form factor $A_3$ 
\begin{equation}
	K_{31}A_3=0,\qquad \qquad\qquad K_{32}A_3=0
\end{equation}
where we have defined 
\begin{equation}
	K_i=\frac{\partial^2}{\partial p_i^2}+\frac{\left(d+1-2 \Delta_i\right)}{p_i} \frac{\partial}{\partial p_i}, \qquad\qquad K_{i j}=K_i-K_j
\end{equation}
Recalling the following property of the 3K integrals
\begin{equation}
	K_{n m} J_{N\left\{k_j\right\}}=-2 k_n J_{N+1\left\{k_j-\delta_{j n}\right\}}+2 k_m J_{N+1\left\{k_j-\delta_{j m}\right\}},
\end{equation}
we can write the most general solution of the primary equations as\\
\begin{equation}
	A_3= \zeta_1 \,J_{\{5,0,0,0\}}
\end{equation}
where $\zeta_1$ is an arbitrary constant. Note that this solution is symmetric under the exchange of momenta $p_1\leftrightarrow p_2$. 
Indeed, from the definition of the 3K integral, it follows that for any permutation $\sigma$ of the set $\{1, 2, 3\}$ we have
\begin{equation}
	J_{N\left\{k_{\sigma(1)} ,k_{\sigma(2)} ,k_{\sigma(3)}\right\}}\left(p_1, p_2, p_3\right)=J_{N \left\{k_1, k_2, k_3\right\}}\left(p_{\sigma^{-1}(1)}, p_{\sigma^{-1}(2)}, p_{\sigma^{-1}(3)}\right)
\end{equation}
However, due to the Bose symmetry, the form factor $A_3$ needs to be antisymmetric under the exchange of momenta $p_1\leftrightarrow p_2$.  This leads to
\begin{equation}
	\zeta_1=0 \quad \Longrightarrow \quad A_3=0
\end{equation}
After setting $A_3=0$, the explicit form of the primary equations involving the form factor $A_4$ can be written as
\begin{equation}
	K_{31}A_4=0,\qquad \qquad\qquad K_{32}A_4=0
\end{equation}
The solution is given by
\begin{equation}
	A_4=\zeta_2 \, J_{\{3,0,0,0\}},
\end{equation}
where $\zeta_2$ is an arbitrary constant.
Once again, due to the Bose symmetry, $A_4$ needs to be antisymmetric under the exchange of momenta $p_1\leftrightarrow p_2$. This leads to
\begin{equation}
	\zeta_2=0 \quad \Longrightarrow \quad  A_4=0.
\end{equation}
After setting $A_4=0$, we can write the remaining primary equations as\footnote{For simplicity, we are actually considering on the right side two equations that are obtained by a combination of primary and secondary equations: $0=C_{21}-C_{33}$ and $0=C_{23}-C_{34}$. The contribution of the anomalous term coming from eq. \eqref{eq:anomcontrscwi} does not appear in such combinations of equations.}
\begin{equation}\label{eq:eqprimariee}
	\begin{aligned}
		&0=K_{31}A_1, \qquad\qquad\qquad
		&&0=K_{32}A_1+\frac{2}{p_1^2} \left(p_1\frac{\partial }{\partial p_1}
		-4 \right) A_1(p_1\leftrightarrow p_2)\\
		&0=K_{31}A_2+4A_1, 
		&&0=K_{32}A_2+\frac{2}{p_1^2} \left(p_1\frac{\partial }{\partial p_1}
		-4 \right) A_2(p_1 \leftrightarrow p_2)+4A_1
	\end{aligned}
\end{equation}
These equations can also be reduced to a set of homogenous equations by repeatedly applying the
operator $K_{ij}$
\begin{equation}
	\begin{aligned}
		&0=K_{31}A_1, \qquad \qquad \qquad &&0=K_{32}K_{32}A_1\\
		&0=K_{31}K_{31}A_2, \qquad \qquad \qquad &&0=K_{32}K_{32}K_{32}A_2.\\
	\end{aligned}
\end{equation}
The most general solution of such homogenous equations can be written in terms of the following 3K integrals
\begin{equation} \label{eq:solomogenee}
	\begin{aligned}
		A_1= &{\eta_1}\,J_{3\{0,0,0\}}+\eta_2\,J_{4\{0,1,0\}},\\
		A_2= &\theta_1\,J_{4\{1,2,0\}}+\theta_2\,J_{3\{0,2,0\}}+\theta_3 \,J_{3\{1,1,0\}}
		\\&\qquad+\theta_4\,J_{2\{0,1,0\}}+\theta_5\,J_{2\{1,0,0\}}+\theta_6\,J_{1\{0,0,0\}}+\theta_7\, J_{3\{0,1,1\}}+\theta_8\,J_{2\{0,0,1\}},
	\end{aligned}
\end{equation}
where $\eta_i$ and $\theta_i$ are arbitrary constants.
The explicit form of such 3K integrals can be determined by following the procedure in \cite{Bzowski:2013sza,Bzowski:2015yxv,Bzowski:2020lip}.
Before moving on, we need to examine the divergences in the 3K integrals.
For a more detailed review of the topic, see Appendix \ref{appendix:3kint} and \cite{Bzowski:2013sza,Bzowski:2015pba,Bzowski:2015yxv}. In general, it can be shown that a 3K integral $I_{\alpha\{\beta_1,\beta_2,\beta_3\}}$ diverges if
\begin{equation}
	\alpha+1 \pm \beta_1 \pm \beta_2 \pm \beta_3=-2 k \quad, \quad k=0,1,2, \dots
\end{equation}
If the above condition is satisfied, we need to regularize the integral. Therefore, we shift its parameters by small amounts proportional to a regulator $\epsilon$ according to the formula
\begin{equation}
	\begin{aligned}
		&I_{\alpha\left\{\beta_1, \beta_2, \beta_3\right\}} &&\longmapsto\quad I_{{\alpha+u\epsilon}\left\{{\beta}_1+v_1\epsilon ,\, {\beta}_2 +v_2\epsilon,\, {\beta}_3+v_3\epsilon \right\}}\\
		&J_{N\left\{k_1 ,k_2, k_3\right\}} &&\longmapsto \quad J_{{N+u\epsilon}\left\{{k}_1 +v_1\epsilon,\,{k}_2+v_2\epsilon, \,{k}_3+v_3\epsilon \right\}}.
	\end{aligned}
\end{equation}
The arbitrary numbers $u$, $v_1$, $v_2$ and $v_3$ specify the direction of the shift. 
In general the regulated integral exists, but exhibits singularities when $\epsilon$ is taken to zero.
If a 3K integral in our solution diverges, we can expand the coefficient in front of such integral in the solution in powers of $\epsilon$ 
\begin{equation}
	\begin{aligned}
		\eta_i=  \sum_{j=-\infty}^\infty \eta_i^{(j)}\epsilon^j\qquad\qquad
		\theta_i=  \sum_{j=-\infty}^\infty \theta_i^{(j)}\epsilon^j,\\
	\end{aligned}
\end{equation}
and then we can require that our entire solution is finite for $\epsilon\rightarrow 0$ by constraining the coefficients $ \eta_i^{(j)}$ and $ \theta_i^{(j)}$. Both of the 3K integrals appearing in the eq$.$ \eqref{eq:solomogenee} in the solution for $A_1$ diverge like $1/\epsilon$. Therefore, we require
\begin{equation}
	\eta_1= \eta_1^{(0)}+\eta_1^{(1)} \epsilon ,\qquad\qquad\qquad
	\eta_2= \eta_2^{(0)}+\eta_2^{(1)} \epsilon. 
\end{equation}
Higher order terms do not contribute to the solution and therefore they can be neglected. In the case of $A_2$, since some of the 3K integrals diverge like $1/\epsilon^2$, we need to set
\begin{equation}
	\begin{gathered}
		\theta_1= \theta_1^{(0)}+\theta_1^{(1)} \epsilon +\theta_1^{(2)} \epsilon ^2,\qquad
		\theta_2= \theta_2^{(0)}+\theta_2^{(1)} \epsilon +\theta_2^{(2)} \epsilon ^2,\qquad
		\theta_3=\theta_3^{(0)}+\theta_3^{(1)} \epsilon +\theta_3^{(2)} \epsilon ^2,\\
		\theta_4= \theta_4^{(0)}+\theta_4^{(1)} \epsilon +\theta_4^{(2)} \epsilon ^2,\qquad
		\theta_5= \theta_5^{(0)}+\theta_5^{(1)} \epsilon +\theta_5^{(2)} \epsilon ^2,\qquad
		\theta_6= \theta_6^{(0)}+\theta_6^{(1)} \epsilon +\theta_6^{(2)} \epsilon ^2,\\
		\theta_7= \theta_7^{(0)}+\theta_7^{(1)} \epsilon ,\qquad\qquad\qquad
		\theta_8= \theta_8^{(0)}+\theta_8^{(1)} \epsilon.
	\end{gathered}
\end{equation}
The last step consists in analyzing all the conformal constraints on the numerical coefficients $\eta^{(j)}_i$ and $\theta^{(j)}_i$. In order to do that, we insert our solution back into the primary non-homogenous equations \eqref{eq:eqprimariee} and into the secondary equations. The explicit form of the secondary equations is given by\footnote{Not of the secondary equations are independent from each other. Here we listed only the relevant ones.}
\small
\begin{equation}
	\begin{aligned}
		&0=
		-2 p_1 \frac{\partial  }{\partial p_1}A_1+2 p_2 \frac{\partial  }{\partial p_2}A_1(p_1\leftrightarrow p_2)\\[5pt]
		&0=-\left(p_1^2-p_2^2+p_3^2\right) A_1+\left(-p_1^2+p_2^2+p_3^2 \right) A_1(p_1\leftrightarrow p_2)-2 p_1 \frac{\partial  }{\partial p_1}A_2+2 p_2 \frac{\partial  }{\partial p_2}A_2(p_1\leftrightarrow p_2)\\&
		\quad +2 A_2-2 A_2(p_1\leftrightarrow p_2)\\[5pt]
		&0=-\frac{2 p_2^3 }{p_3^2}\frac{\partial  }{\partial p_2}A_1(p_1\leftrightarrow p_2)-2 \left(\frac{ p_2^2+p_3^2 }{p_3^2}\right)p_2\frac{\partial  }{\partial p_2}A_1+ \left(-\frac{2 p_2^2}{p_3^2}+\frac{p_3^2-p_2^2-p_1^2}{p_1^2}\right) p_1 \frac{\partial  }{\partial p_1}A_1\\&\quad+2 p_2^2\left(\frac{p_3^2-p_1^2}{p_3^2 p_1}\right) \frac{\partial  }{\partial p_1}A_1(p_1\leftrightarrow p_2)
		-4 p_2^2 \left(\frac{2}{p_1^2}+\frac{1}{p_3^2}\right)A_1(p_1\leftrightarrow p_2)
		+4 \left(\frac{p_1^2+p_2^2-p_3^2}{p_1^2} -\frac{p_2^2}{p_3^2}\right)A_1
		\\&\quad-\frac{2 }{p_1}\frac{\partial  }{\partial p_1}A_2+\frac{8}{p_1^2} A_2-\frac{64 i a p_2^2}{p_3^2}
		\\[5pt]&
		0=
		-\left(\frac{p_1^2+p_2^2-p_3^2}{p_3^2}\right)p_1  \frac{\partial  }{\partial p_1}A_1
		-\frac{\left(p_1^2-2 p_3^2 \right) \left(p_1^2+p_2^2-p_3^2 \right) }{p_1 p_3^2}\frac{\partial  }{\partial p_1}A_1(p_1\leftrightarrow p_2)-\left(\frac{ p_1^2+p_2^2-p_3^2}{p_3^2}\right)p_2 \frac{\partial  }{\partial p_2}A_1\\&\quad
		-\left(\frac{p_1^2+p_2^2-3 p_3^2}{p_3^2} \right)p_2 \frac{\partial  }{\partial p_2}A_1(p_1\leftrightarrow p_2)
		-2\left(\frac{p_1^2+p_2^2-2p_3^2}{p_3^2}+4 \, \frac{p_1^2+p_2^2-p_3^2}{p_1^2}\right) A_1(p_1\leftrightarrow p_2)\\&\quad
		-2 \left(\frac{ p_1^2+p_2^2-2 p_3^2}{p_3^2}\right) A_1+\frac{2 }{p_1}\frac{\partial  }{\partial p_1}A_2(p_1\leftrightarrow p_2)-\frac{8 }{p_1^2}A_2(p_1\leftrightarrow p_2)-\frac{32 i a \left(p_1^2+p_2^2-p_3^2\right)}{p_3^2}
		\\[5pt]&
		0=
		\frac{2 p_1 }{p_3^2}\frac{\partial  }{\partial p_1}A_1 +2 \left(\frac{p_1^2-p_3^2 }{p_3^2\, p_1}\right) \frac{\partial  }{\partial p_1}A_1(p_1\leftrightarrow p_2)+\frac{2 p_2 }{p_3^2}\frac{\partial  }{\partial p_2}A_1+\frac{2 p_2 }{p_3^2}\frac{\partial  }{\partial p_2}A_1(p_1\leftrightarrow p_2)\\&\quad
		+4\left( \frac{2}{p_1^2}+\frac{1}{p_3^2} \right)A_1(p_1\leftrightarrow p_2)+\frac{4 }{p_3^2}A_1+\frac{64 i a}{p_3^2}
		\\[5pt]&
		0=
		-\frac{2 p_1 }{p_3^2}\frac{\partial  }{\partial p_1}A_2
		+2\left(\frac{p_3^2-p_1^2}{p_3^2\,p_1} \right)\frac{\partial  }{\partial p_1}A_2(p_1\leftrightarrow p_2)
		-\frac{2 p_2 }{p_3^2}\frac{\partial  }{\partial p_2}A_2-\frac{2 p_2 }{p_3^2}\frac{\partial  }{\partial p_2}A_2(p_1\leftrightarrow p_2)-\frac{8 }{p_1^2}A_2(p_1\leftrightarrow p_2)\\&\quad
		+\frac{32 i a \left(p_1^2+p_2^2-p_3^2\right)}{p_3^2}
	\end{aligned}
\end{equation}
\normalsize
We can solve all these equations by performing the limit $p_i \rightarrow 0$, as explained in the Appendix \ref{appzerolimit3k}.
After some lengthy computations, using all the properties of the 3K integral listed in the Appendix \ref{appendix:3kint}, we find that all the non-vanishing coefficients $\eta^{(j)}_i$ and $\theta^{(j)}_i$ depend on the anomaly coefficient $a_2$ of eq$.$ \eqref{eq:anomaliachirale}. In particular the final solution can be written in the compact form
\begin{equation}\label{eq:resultffconformi}
	\begin{aligned}
		&A_1=-4\, i \, a_2\, p_2^2\, I_{5\{2,1,1\}}\\
		&A_2=-8\, i \, a_2\, p_2^2\, \bigg(p_3^2\, I_{4\{2,1,0\}}-1 \bigg)\\
		&A_3=0\\
		&A_4=0.\
	\end{aligned}
\end{equation}

\section{Perturbative realization} \label{pertrealiz}
In this section we compute the $\langle TTJ_5\rangle$ correlator perturbatevely at one-loop, working in the Breitenlohner-Maison scheme.
For this analysis we shift to the Minkowski space where
\be
e^{i \cS[g]} \equiv \int [d \Phi]\, e^{i {{S}_{0}}[\Phi,\, g]}
\ee
We consider the following action with a fermionic field in a gravitational and axial gauge field background
\begin{equation}
	S_0=\int d^d x \, \frac{e}{2}\,  e_a^\mu \bigg[{i}  \bar{\psi} \gamma^a\left(D_\mu \psi\right)-{i} \left(D_\mu \bar{\psi}\right) \gamma^a \psi\bigg]
\end{equation}
where $e_a^\mu $ is the vielbein, $e$ is its determinant and $D_\mu$ is the covariant derivative defined as
\begin{equation}
	\begin{aligned}
		& D_\mu \psi=\left(\nabla_\mu+i g \gamma_5 A_\mu\right) \psi=\left(\partial_\mu+i g \gamma_5A_\mu+\frac{1}{2} \omega_{\mu a b} \Sigma^{a b}\right) \psi, \\
		& D_\mu \bar{\psi}=\left(\nabla_\mu-i g \gamma_5 A_\mu\right) \bar{\psi}=\left(\partial_\mu-i g\gamma_5 A_\mu-\frac{1}{2} \omega_{\mu a b} \Sigma^{a b}\right) \bar{\psi}.
	\end{aligned}
\end{equation}
$\Sigma^{ab}$ are the generators of the Lorentz group in the case of a spin 1/2-field, while the spin connection is given by
\begin{equation}
	\omega_{\mu a b} \equiv e_a^\nu\left(\partial_\mu e_{\nu b}-\Gamma_{\mu \nu}^\lambda e_{\lambda b}\right).
\end{equation}
The Latin and Greek indices are related to the (locally) flat basis and the curved background respectively. Using the explicit expression of the generators of the Lorentz group one can re-express the action $S_0$ as follows
\begin{equation}
	S_0=\int d^d x \, e \left[\frac{i}{2} \bar{\psi} e_a^\mu \gamma^a\left(\partial_\mu \psi\right)-\frac{i}{2}\left(\partial_\mu \bar{\psi}\right) e_a^\mu \gamma^a \psi-g A_\mu  \bar{\psi}  e_a^\mu  \gamma^a\gamma_5\psi  +\frac{i}{4} \omega_{\mu a b} e_c^\mu \bar{\psi} \gamma^{a b c} \psi\right]
\end{equation}
with
\begin{equation}
	\gamma^{abc}=\{\Sigma^{ab},\gamma^c \}.
\end{equation}
Taking a first variation of the action with respect to the metric one can construct the energy momentum tensor as
\begin{equation}
	T^{\mu \nu}=-\frac{i}{2}\left[\bar{\psi} \gamma^{(\mu} \nabla^{\nu)} \psi-\nabla^{(\mu} \bar{\psi} \gamma^{\nu)} \psi-g^{\mu \nu}\left(\bar{\psi} \gamma^\lambda \nabla_\lambda \psi-\nabla_\lambda \bar{\psi} \gamma^\lambda \psi\right)\right]-g \bar{\psi}\left(g^{\mu \nu} \gamma^\lambda A_\lambda-\gamma^{(\mu} A^{\nu)}\right) \gamma_5 \, \psi.
\end{equation}
The computation of the vertices can be done by taking functional derivatives of the action with respect to the metric and the gauge field and Fourier transforming to momentum space. Their explicit expressions is reported in Appendix \ref{appvertices}.

\subsection{Feynman diagrams}
\begin{figure}[h]
	\includegraphics[scale=0.5]{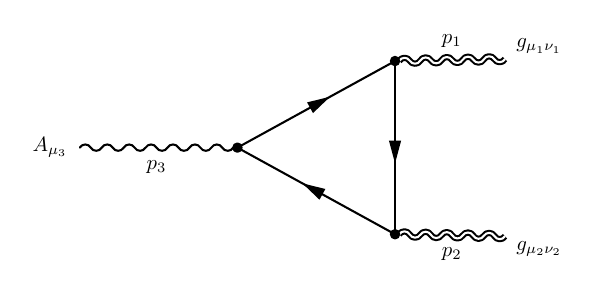}
	\includegraphics[scale=0.5]{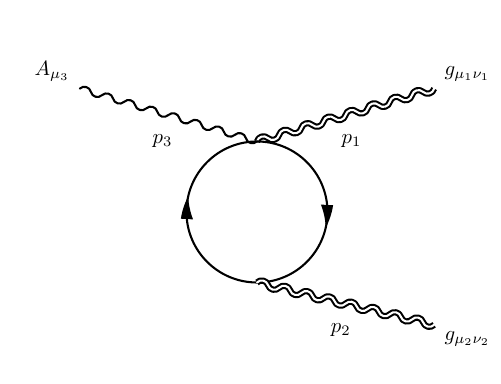}
	\includegraphics[scale=0.5]{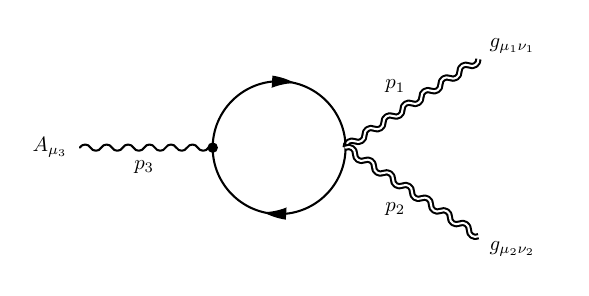}
	\includegraphics[scale=0.5]{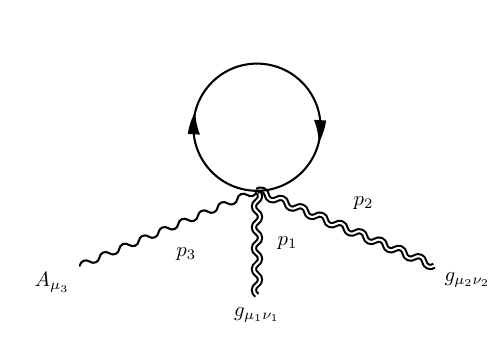}
	\caption{Feynman Diagrams of the three different topologies appearing in the perturbative computation.} 
	\label{fig:feynmdiagr}
\end{figure}
The $\langle  TTJ_5\rangle$ correlator around flat space is extracted by taking three functional derivatives of the  effective action with respect to the metric and the gauge field, evaluated when the sources are turned off
\begin{equation}
	\left\langle T^{\mu_1 \nu_1}\left({x}_{1}\right) T^{\mu_2 \nu_2}\left({x}_{2}\right) J_5^{\mu_3 }\left({x_3}\right)\right\rangle\equiv 
	\left.4 \, \frac{\delta^3 \cS}{\d g_{\mu_1 \nu_1}(x_1)\d g_{\mu_2 \nu_2}(x_2)\d A_{\mu_3}(x_3)}\right|_{\scriptsize{\begin{aligned}
				{g=\eta;}\\{A=0;}
	\end{aligned}}}
\end{equation}
Having denoted with $S_0$ the conformal invariant classical action, recalling eq$.$ \eqref{Sexact}, we can write
\begin{equation}
	\begin{aligned}
		\langle &T^{\mu_1 \nu_1}\left({x}_{1}\right) T^{\mu_2 \nu_2}\left({x}_{2}\right) J_5^{\mu_3 }\left({x_3}\right)\rangle=\\ &
		\,\,\, 4 \, \bigg\{  
		- i \left\langle \frac{\delta S_0}{\delta g_1 }\frac{\delta S_0}{\delta g_2}\frac{\delta S_0}{\delta A_3}\right\rangle
		- \left\langle \frac{\delta^2 S_0}{\delta g_1 \delta g_2}\frac{\delta S_0}{\delta A_3} \right\rangle
		- \left\langle \frac{\delta^2 S_0}{\delta g_1 \delta A_3}\frac{\delta S_0}{\delta g_2} \right\rangle
		- \left\langle \frac{\delta^2 S_0}{\delta g_2 \delta A_3}\frac{\delta S_0}{\delta g_1} \right\rangle 
		+ i \left\langle \frac{\delta^3 S_0}{\delta g_1 \delta g_2 \delta A_3} \right\rangle
		\bigg\}
	\end{aligned}
\end{equation}
where for the sake of simplicity we have used the notation $g_i = g_{\mu_i\nu_i}(x_i)$ and $A_i = A_{\mu_i}(x_i)$. The angle brackets denote the vacuum expectation value and each of the terms correspond to a  Feynman diagram of specific topology. In particular, the first term has a triangle topology while the others are all bubble diagrams, except for the last one, which is a tadpole (see Fig$.$ \ref{fig:feynmdiagr}).
The contribution of the triangle diagrams is given by 
\begin{equation}
	\begin{aligned}
		V_{1}^{\mu_1\nu_1\mu_2\nu_2\mu_3}=-i^3\int \frac{d^d l}{(2\pi)^d}\,\, \frac{\text{tr}\big[V^{\mu_1\nu_1}_{g \bar{\psi}\psi}(l-p_1,l)\, (\cancel{l}-\cancel{p}_1)   V^{\mu_3}_{A \bar{\psi}\psi}
			(\cancel{l}+\cancel{p}_2)
			V^{\mu_2\nu_2}_{g \bar{\psi}\psi}(l,l+p_2)
			\cancel{l}
			\big]}{(l-p_1)^2\,(l+p_2)^2\, l^2} +\textrm{exchange}
	\end{aligned}
\end{equation}
while the bubble diagrams are
\begin{equation}
	\begin{aligned}
		V_2^{\mu_1\nu_1\mu_2\nu_2\mu_3}=-i^2\int \frac{d^d l}{(2\pi)^d}\,\, \frac{\text{tr}\big[V^{\mu_1\nu_1\mu_3}_{gA \bar{\psi}\psi}\, (\cancel{l}+\cancel{p}_2)   V^{\mu_2\nu_2}_{g \bar{\psi}\psi}
			(l,l+p_2) \cancel{l}
			\big]}{(l+p_2)^2\,l^2} +\textrm{exchange}
	\end{aligned}
\end{equation}
and
\begin{equation}
	\begin{aligned}
		V_3^{\mu_1\nu_1\mu_2\nu_2\mu_3}=-i^2\int \frac{d^d l}{(2\pi)^d}\,\, \frac{\text{tr}\big[V^{\mu_1\nu_1\mu_2\nu_2}_{gg \bar{\psi}\psi}(p_1,p_2,l-p_1-p_2,l)\, (\cancel{l}-\cancel{p}_1-\cancel{p}_2)   V^{\mu_3}_{A \bar{\psi}\psi}  \cancel{l}
			\big]}{(l-p_1-p_2)^2\,l^2}. 
	\end{aligned}
\end{equation}
After performing the integration, one can verify  that $V_2^{\mu_1\nu_1\mu_2\nu_2\mu_3}$ vanishes.
Lastly, the tadpole diagram is given by
\begin{equation}
	\begin{aligned}
		V_4^{\mu_1\nu_1\mu_2\nu_2\mu_3}=-i\int \frac{d^d l}{(2\pi)^d}\,\, \frac{\text{tr}\big[V^{\mu_1\nu_1\mu_2\nu_2\mu_3}_{ggA \bar{\psi}\psi}\, \cancel{l}  
			\big]}{l^2}. 
	\end{aligned}
\end{equation}
This last diagram vanishes since it contains the trace of two $\gamma$'s and a $\gamma_5$.
The perturbative realization of the correlator will be written down as the sum of the amplitudes, formally given by the expression
\begin{equation}
	\left\langle T^{\mu_{1} \n_{1}} T^{\mu_{2} \n_2} J_5^{\mu_{3}}\right\rangle=4\, \sum_{i=1}^4\, V_{i}^{\mu_1\nu_1\mu_2\nu_2\mu_3}.
\end{equation}

\subsection{Reconstruction of the correlator}

The perturbative realization of the $\langle TTJ_5\rangle $ satisfies the (anomalous) conservation and trace WIs. Therefore, the correlator can be decomposed as described in section \ref{sectiondecompcorr}. In particular, it is comprised of two terms
\begin{equation}
	\left\langle T^{\mu_{1} \n_{1}} T^{\mu_{2} \n_2} J_5^{\mu_{3}}\right\rangle=\left\langle t^{\mu_{1} \n_{1}} t^{\mu_{2}\n_2} j_5^{\mu_{3}}\right\rangle+\left\langle t^{\mu_{1} \n_{1}} t^{\mu_{2}\n_2} j_{5\, l o c}^{\mu_{3}}\right\rangle.
\end{equation}
The anomalous pole is given by
\begin{equation}
	\left\langle t^{\mu_{1} \n_{1}} t^{\mu_{2}\n_2} j_{5\, l o c}^{\mu_{3}}\right\rangle= \frac{g}{96\pi^2}\,  \frac{p_3^{\mu_3}}{p_3^2} \, (p_1 \cdot p_2) \left\{ \left[\varepsilon^{\nu_1 \nu_2 p_1 p_2}\left(g^{\mu_1 \mu_2}- \frac{p_1^{\mu_2} p_2^{\mu_1}}{p_1 \cdot p_2}\right) +\left( \mu_1 \leftrightarrow \nu_1 \right) \right] +\left( \mu_2 \leftrightarrow \nu_2 \right) \right\},
\end{equation}
which corresponds to eq$.$ \eqref{eq:anompolettj} with 
\begin{equation} \label{eq:coeffperturbativo}
	a_2= - \frac{i g}{384 \pi^2}.
\end{equation}
The transverse-traceless part $\left\langle t^{\mu_{1} \n_{1}} t^{\mu_{2}\n_2} j_5^{\mu_{3}}\right\rangle$ can be expressed in terms of four form factors as described in eq$.$ \eqref{eq:decomptt}. The perturbative calculation in four dimensions gives
\begin{equation} \label{eq:pertresultff}
	\begin{aligned}
		A_1&=\frac{g p_2^2}{24 \pi^2 \lambda^4}
		\Bigg\{A_{11} 
		+A_{12} \log\left(\frac{p_1^2}{p_2^2}\right)+A_{13} \log\left(\frac{p_1^2}{p_3^2}\right)
		+A_{14} \, \,C_0(p_1^2,p_2^2,p_3^2)
		\Bigg\}, \\[8pt]
		A_2&=\frac{gp_2^2}{48\pi^2\lambda^3}\Bigg\{
		A_{21}
		+A_{22} \log\left(\frac{p_1^2}{p_2^2}\right)+A_{23} \log\left(\frac{p_1^2}{p_3^2}\right)
		+A_{24} \, \, C_0(p_1^2,p_2^2,p_3^2)
		\Bigg\}, \\[8pt]
		A_3&=0,\\[8pt]
		A_4&=0\\[8pt],
	\end{aligned}
\end{equation}
where $C_0$ in Minkowski space is the master integral 
\begin{equation}
	C_{0}(p_1^2,p_2^2,p_3^2) \equiv \frac{1}{ i \pi^2}  \int d^d l \frac{1}{l^2(l-p_1)^2(l+p_2)^2}
\end{equation}
and we have introduced the following quantities
\footnotesize
\begin{equation}
	\begin{aligned}
		A_{11}=&- \lambda \Bigg[2 p_1^{10}-p_1^8 (p_2^2+p_3^2)-2 p_1^6 (5 p_2^4-48 p_2^2 p_3^2+5 p_3^4)+4 p_1^4 (p_2^2+p_3^2) (4 p_2^4-23 p_2^2 p_3^2+4 p_3^4)\\&\qquad
		-8 p_1^2 (p_2^2-p_3^2)^2 (p_2^4+4 p_2^2 p_3^2+p_3^4)
		+(p_2^2-p_3^2)^4 (p_2^2+p_3^2)\Bigg]		\\
		A_{12}=&+2 p_2^2 \Bigg[p_3^2 (p_3^2-p_2^2)^5+p_1^{10} (38 p_3^2-12 p_2^2)+p_1^8 (18 p_2^4+41 p_2^2 p_3^2-121 p_3^4)
		-4 p_1^6 (3 p_2^6+46 p_2^4 p_3^2-38 p_2^2 p_3^4-26 p_3^6)\\&\qquad
		+p_1^4 (p_2-p_3) (p_2+p_3) (3 p_2^6+95 p_2^4 p_3^2+215 p_2^2 p_3^4+11 p_3^6)+14 p_1^2 p_3^2 (p_2^2-p_3^2)^3 (p_2^2+p_3^2)+3 p_1^{12}\Bigg]\\
		A_{13}=&+2 p_3^2 \Bigg[3 p_1^{12}+2 p_1^{10} (19 p_2^2-6 p_3^2)+p_1^8 (-121 p_2^4+41 p_2^2 p_3^2+18 p_3^4)+4 p_1^6 (26 p_2^6+38 p_2^4 p_3^2-46 p_2^2 p_3^4-3 p_3^6)\\&\qquad
		-14 p_1^2 p_2^2 (p_2^2-p_3^2)^3 (p_2^2+p_3^2)-p_1^4 (p_2-p_3) (p_2+p_3) (11 p_2^6+215 p_2^4 p_3^2+95 p_2^2 p_3^4+3 p_3^6)+p_2^2 (p_2^2-p_3^2)^5\Bigg]\\
		A_{14}=&-24 p_1^4 p_2^2 p_3^2 \Bigg[ (p_1^2-p_2^2)^3 (2 p_1^2+3 p_2^2)
		-3 p_3^4 (p_1^4+4 p_1^2 p_2^2-4 p_2^4)-3 p_3^2 (p_1^6-6 p_1^4 p_2^2+4 p_1^2 p_2^4+p_2^6)-3 p_3^8
		p_3^6 (7 p_1^2-3 p_2^2)\Bigg]\\
		A_{21}=&-\lambda \Bigg[2 p_3^6 (3 p_1^2+p_2^2)+4 p_1^2 p_3^4 (3 p_2^2-2 p_1^2)+(p_1^2-p_2^2)^4+2 p_3^2 (p_1-p_2) (p_1+p_2) (p_1^4+8 p_1^2 p_2^2+p_2^4)-p_3^8\Bigg]\\
		A_{22}=&-2 p_2^2 p_3^2 \Bigg[-17 p_1^8+p_1^6 (28 p_2^2+26 p_3^2)-4 p_1^4 (p_2^4+15 p_2^2 p_3^2)+(p_2^2-p_3^2)^4
		-2 p_1^2 (p_2^2-p_3^2)^2 (4 p_2^2+5 p_3^2)\Bigg] \\
		A_{23}=&+2 p_3^2 \Bigg[2 p_1^{10}-p_1^8 (p_2^2+6 p_3^2)+p_1^6 (-10 p_2^4+46 p_2^2 p_3^2+6 p_3^4)-2 p_1^2 (4 p_2^2+5 p_3^2) (p_2^3-p_2 p_3^2)^2+p_2^2 (p_2^2-p_3^2)^4
		\\&\qquad
		+2 p_1^4 (8 p_2^6-21 p_2^4 p_3^2-18 p_2^2 p_3^4-p_3^6)
		\Bigg]\\
		A_{24}=&-12 p_1^4 p_2^2 p_3^2 \Bigg[ p_3^4 (3 p_2^2-5 p_1^2)+(p_1^2-p_2^2)^3+p_3^2 (p_1^4+4 p_1^2 p_2^2-5 p_2^4)+3 p_3^6\Bigg] 
	\end{aligned}
\end{equation}
\normalsize
with the K\"{a}llen $\lambda$-function given by
\begin{equation}
	\lambda \equiv \lambda(p_1,p_2,p_3)=\left(p_1-p_2-p_3\right)\left(p_1+p_2-p_3\right)\left(p_1-p_2+p_3\right).\left(p_1+p_2+p_3\right)
	\label{Kallen}
\end{equation}

\section{Matching the perturbative solution} \label{matchpert}
In this section, we verify the matching between the perturbative form factors in Eq. \eqref{eq:pertresultff} and the non-perturbative ones in Eq. \eqref{eq:resultffconformi}. First of all, one can immediately see that $A_3$ and $A_4$ vanish in both calculations. On the other hand, in order to verify the matching between the first two form factors, we will need to rewrite the 3K integrals in the conformal solution in terms of the master integral $C_0$. For this purpose, we recall the reduction relations presented in \cite{Bzowski:2015yxv, Bzowski:2020lip}
\begin{equation}
	I_{\alpha\left\{\beta_1 \beta_2 \beta_3\right\}}=(-1)^{\beta_t} \mathrm{~K}_{j, \beta_j}^{\left|n_0\right|-1}\left[p_1^{2 \beta_1} p_2^{2 \beta_2} p_3^{2 \beta_3}\left(\frac{1}{p_1} \frac{\partial}{\partial p_1}\right)^{\beta_1}\left(\frac{1}{p_2} \frac{\partial}{\partial p_2}\right)^{\beta_2}\left(\frac{1}{p_3} \frac{\partial}{\partial p_3}\right)^{\beta_3} I_{1\{000\}}\right]
\end{equation}
Moreover, the integral $I_{1 \{{0,0,0}\}}$ is related to the massless scalar $1$-loop $3$-point momentum-space integral
\begin{align}
	I_{1\{0,0,0\}}=(2\pi)^2K_{4,\{1,1,1\}}=(2\pi)^2\int\frac{d^4k}{(2\pi)^4}\frac{1}{k^2\,(k-p_1)^2\,(k+p_2)^2}=\frac{1}{4}C_{0}(p_1^2,p_2^2,p_3^2)
\end{align}
where 
\begin{align}
	K_{d\{\delta_1\delta_2\delta_3\}}\equiv \int\frac{d^dk}{(2\pi)^d}\frac{1}{(k^2)^{\delta_3}\,((k-p_1)^2)^{\delta_2}\,((k+p_2)^2)^{\delta_1}}.
\end{align}
Hence, it follows that the 3K integrals in our conformal solutions \eqref{eq:resultffconformi} are finite and can be reduced to 
\begin{equation}
	\begin{aligned}
		& I_{5\{2,1,1\}}=\frac{i}{4} p_1 p_2 p_3 \left(p_1 \frac{\partial}{\partial p_1}-1\right) \frac{\partial^3}{\partial p_1 \partial p_2 \partial p_3} C_0(p_1^2,p_2^2,p_3^2)\\
		& I_{4\{2,1,0\}}=-\frac{i}{4}p_1  p_2\left( p_1\frac{\partial}{\partial p_1}-1\right)\frac{\partial^2}{\partial p_1 \partial p_2 } C_0(p_1^2,p_2^2,p_3^2)
	\end{aligned}
	\label{eqc}
\end{equation}
By using the relations of the derivative acting on the master integral in Appendix \ref{appmasterintegral} and setting the anomalous coefficient as in eq$.$ \eqref{eq:coeffperturbativo}, one can then verify the matching between the perturbative and non-perturbative form factors.

\section{The anomaly pole of the gravitational anomaly and the sum rules}\label{apolesumrule}
As we have already mentioned in the previous sections, it is clear that our result does not depend on the specific expression of the current $J_5$ appearing in the correlator, since we have been using only the general symmetry properties of this 3-point function and its anomaly content in order to solve the conformal constraints. \\
Being the result unique and expressed in terms of a single constant, it shows that in a parity-odd CFT the gravitational anomaly vertex is generated by the exchange of an anomaly pole, with the entire correlator built around such massless pole and the value of its residue.  
Since this massless exchange was also present in the perturbative analysis of \cite{Dolgov:1988qx},  we are now going to elaborate on those previous findings under the light of our current result. \\
\subsection{Duality symmetry}
The Maxwell equations in the absence of charges and currents satisfy the duality symmetry 
($E\to B$ and $B\to -E$). The symmetry can be viewed as a special case of a continuous symmetry

\beq
\delta F^{\mu\nu}_V=\beta \tilde{F}^{\mu\nu}_V
\eeq
where $\delta\beta$ is an infinitesimal $SO(2)$ rotation and $\tilde{F}^{\mu\nu}= \epsilon^{\mu\nu\rho\sigma}F_{\rho\sigma}/2$. Its finite form 

\begin{equation}
\begin{pmatrix}
 E\\
 B \\
\end{pmatrix}=
\begin{pmatrix}
 \cos\beta&\sin\beta\\
 -\sin\beta&\cos\beta \\
\end{pmatrix}\begin{pmatrix}
E\\B\\
\end{pmatrix}
\end{equation}
is indeed a symmetry of the equations of motion, but not of the Maxwell action.  
Notice that the action
\beq
\mathcal{S}=\int d^4 x F_V^{\mu\nu} F_{V\, \mu\nu}^{}
\eeq
 is invariant under an infinitesmal transformation modulo a total derivative. For $\beta=\pi/2$, the discrete case, then the action flips sign since $(F^2_V\to - \tilde{F}^2_V)$. \\
In general, the infinitesimal variation of the action takes the form 
\beq
\label{sym}
\delta_\beta {S}_{0}=-\beta \int d^4 x \partial_\mu \left(\tilde{F}^{\mu\nu}_V \,V_\nu\right).
\eeq
Due to the equivalence (dual Bianchi identity)
\beq
\partial_\nu F_V^{\mu\nu}=0 \leftrightarrow \epsilon^{\mu\nu\rho\sigma}\partial_\nu \tilde{F}_{V\rho\sigma}=0,
\eeq
we can introduce the dual gauge field $\tilde{V}^\mu$ 
\beq
\tilde{F}^{\mu\nu}_V=\partial^\mu \tilde{V}^\nu - \partial^\nu \tilde{V}^\mu,
\eeq
which is related to the original $V_\mu$ one by 
\beq
\partial^\mu \tilde{V}^\nu - \partial^\nu \tilde{V}^\mu=\epsilon^{\mu\nu\rho\sigma}\partial_{\rho}V_\sigma.
\eeq
The current corresponding to the infinitesimal symmetry \eqref{sym} can be expressed in the form
\beq
J^\mu=\tilde{F}_V^{\mu\nu}V_\nu - F_V^{\mu\nu}\tilde{V}_\nu, 
\eeq
whose conserved charge is gauge invariant 
\beq
Q_5=\int d^3 x \left( V\cdot \nabla  \times V -  \tilde{V}\cdot\nabla \times \tilde{V}\right)
\label{kk}
\eeq
after an integration by parts. 
Notice that the two terms on the equation above count the linking number of magnetic and electric lines respectively. In fluid mechanics, helicity is the volume integral of the scalar product of the velocity field with its curl given by
\beq
\mathcal{H}_{fluid}=\int d^3 x \, \vec{v}\cdot \nabla \times \vec{v}
\eeq
and one recognizes in \eqref{kk} the expression 
\beq
\label{qq5}
Q_5=\int d^3 x \left( B\cdot V  - E\cdot \tilde{V}\right)
\eeq
with $B=\nabla\times V$ and $E=-\nabla\times \tilde{V}$, that coincides with the optical helicity of the electromagnetic field \cite{delRio:2020cmv}. \\
 As already mentioned, a perturbative analysis of $\langle TTJ_{CS}\rangle$ has been presented long ago in \cite{Dolgov:1987yp}. \\
The presence of anomaly poles in this correlator can indeed be extracted from \cite{Dolgov:1987yp}, in agreement with our result. Indeed, for on-shell gravitons ($g$) and photons ($\gamma$), the authors obtain, with the inclusion of mass effects in the $\langle AVV\rangle $, $\langle TTJ_{f}\rangle $ and $\langle TT J_{CS}\rangle$ the following expressions for the matrix elements
\beq
\langle 0|J^\mu_f |\gamma \gamma\rangle = f_1(q^2)\frac{q^\mu}{q^2} F_{V\kappa\la}\tilde{F}_V^{\kappa\lambda} 
\eeq
\beq
\langle 0|J^\mu_f | g g\rangle = f_2(q^2)\frac{q^\mu}{q^2} R_{\kappa\la\rho\sigma}\tilde{R}^{\kappa\lambda\rho\sigma} 
\label{oone}
\eeq
\beq
\langle 0|J^\mu_{CS} | g g\rangle = f_3(q^2)\frac{q^\mu}{q^2} R_{\kappa\la\rho\sigma}\tilde{R}^{\kappa\lambda\rho\sigma},
\label{oone2}
\eeq
where $q$ is the momentum of the chiral current. 
The anomaly poles are extracted by including a mass $m$ in the propagators of the loop corrections, in the form of either a fermion mass for the $\langle AVV\rangle$ and the $\langle TTJ_f\rangle $, or working with a 
Proca spin-1 in the case of $\langle TTJ_{CS} \rangle$, and then taking the limit for $m\to 0$.
A dispersive analysis gives for the corresponding spectral densities \cite{Dolgov:1987yp}
\beqa
\Delta_{AVV}(q^2,m)\equiv \textrm{Im} f_1(q^2)&=& \frac{ d_{AVV}}{ q^2}(1- v^2) \log\frac{1 +v }{1-v}\nn\\
\Delta_{TT{J_f}}(q^2,m)\equiv \textrm{Im} f_2(q^2)&=&\frac{d_{TT{J_f}}}{ q^2}(1- v^2)^2 \log\frac{1 +v }{1-v}\nn\\
\Delta_{TTJ_{CS}}(q^2,m)\equiv \textrm{Im} f_3(q^2)&=&\frac{d_{TTJ_{CS}}}{ q^2}v^2(1- v^2)^2 \log\frac{1 +v }{1-v},\nn\\
\eeqa
with $v=\sqrt{1-4 m^2/q^2}$ and $d_{AVV}=-1/2\, \alpha_{em}$, $d_{TT{J_f}}=1/(192 \pi)$ and $d_{TT J_{CS}}=1/(96 \pi)$ being the corresponding anomaly coefficients in the normalization of the currents of \cite{Dolgov:1987yp}, with $\alpha_{em}$ the electromagnetic coupling. \\
Notice the different functional forms of $\Delta_{TT{J_f}}(q^2,m)$ and $\Delta_{TTJ_{CS}}(q^2,m)$ away from the conformal limit, when the mass $m$ is nonzero.
One can easily check that in the massless limit the branch cut present in the previous spectral densities at $q^2= 4m^2$ turns into a pole 
\beq
\lim_{m\to 0}\Delta(q^2,m)\propto \delta(q^2) 
\eeq
in all the three cases. Beside, one can easily show that the same spectral densities satisfy three sum rules 
\beqa 
\int_{4 m^2}^{\infty} ds\,\Delta_{AVV}(s,m)&=& 2 \, d_{AVV}\\
\int_{4 m^2}^{\infty}ds\,\Delta_{TTJ_f}(s,m)&=&\frac{2}{3}\,  d_{TT{J_f}}\\
\int_{4 m^2}^{\infty}ds\,\Delta_{TTJ_{CS}}(s,m)&=& \frac{14}{45} \, d_{TTJ_{CS}},
\eeqa
indicating that for any deformation $m$ from the conformal limit, the integral under $\Delta(s,m)$ is mass independent. Therefore, the numerical value of the area equals the value of the anomaly coefficient in each case.\\
One can verify, from equation \eqref{eq:resultffconformi}, by taking the on-shell photon/graviton limit, that the transverse sector of $\langle TTJ_5\rangle $, corresponding to the form factors $A_1$ and $A_2$, vanishes, since these two form factors are zero, limiting each of these matrix elements to only single form factors, as indicated in \eqref{oone} and \eqref{oone2}.
Then it is clear that, in general, the structure of the anomaly action responsible for the generation of the gravitational chiral anomaly can be expressed in the form
\beq
\mathcal{S}_{anom}\sim \int d^4 x \, d^4 y\, \partial_\lambda A^\lambda\, \frac{1}{\Box}(x,y) R\tilde{R}(y) +\ldots
\eeq 
where the ellipses stand for the transverse sector, and $A_\lambda$ is a spin-1 external source. For on-shell gravitons, as remarked above, this action summarizes the effect of the entire chiral gravitational anomaly vertex, being exactly given by the exchange of a single anomaly pole.

\section{Summary of the results} \label{sumresultsec}
Before coming to our comments and conclusions, for the reader's convenience, we briefly summarize our findings.\\
We have shown that in a general CFT the  $\langle TTJ_5\rangle $ correlator can be written as a sum of two terms
\begin{equation}
	\left\langle T^{\mu_{1} \n_{1}} T^{\mu_{2} \n_2} J_5^{\mu_{3}}\right\rangle=\left\langle t^{\mu_{1} \n_{1}} t^{\mu_{2}\n_2} j_5^{\mu_{3}}\right\rangle+\left\langle t^{\mu_{1} \n_{1}} t^{\mu_{2}\n_2} j_{5\, l o c}^{\mu_{3}}\right\rangle, 
	\end{equation}
the first term being the transverse component and the second, the longitudinal one, expressed in terms of a single anomaly form factor and tensor structure. This is characterized by an interpolating anomaly pole. \\
The anomaly part is given by the expression
\begin{equation}
\label{eew}
	\left\langle t^{\mu_{1} \n_{1}} t^{\mu_{2}\n_2} j_{5\, l o c}^{\mu_{3}}\right\rangle= 4 i a_2 \frac{p_3^{\mu_3}}{p_3^2} \, (p_1 \cdot p_2) \left\{ \left[\varepsilon^{\nu_1 \nu_2 p_1 p_2}\left(g^{\mu_1 \mu_2}- \frac{p_1^{\mu_2} p_2^{\mu_1}}{p_1 \cdot p_2}\right) +\left( \mu_1 \leftrightarrow \nu_1 \right) \right] +\left( \mu_2 \leftrightarrow \nu_2 \right) \right\}
\end{equation}
while the transverse-traceless part is 
\small
\begin{equation}
	\begin{aligned}
		\langle t^{\mu_{1} \nu_{1}}&\left({p}_{1}\right) t^{\mu_{2} \nu_{2}}\left({p}_{2}\right) j_5^{\mu_{3} }\left({p_3}\right)\rangle=\Pi_{\alpha_{1} \beta_{1}}^{\mu_{1} \nu_{1}}\left({p}_{1}\right) \Pi_{\alpha_{2} \beta_{2}}^{\mu_{2} \nu_{2}}\left({p}_{2}\right) \pi_{\alpha_{3}}^{\mu_{3}}\left({p_3}\right) \bigg[
		\\
		&A_1\varepsilon^{p_1\alpha_1\alpha_2\alpha_3}p_2^{\beta_1}p_3^{\beta_2}
		-A_1(p_1\leftrightarrow p_2) \varepsilon^{p_2\alpha_1\alpha_2\alpha_3}p_2^{\beta_1}p_3^{\beta_2}
		+A_2\varepsilon^{p_1\alpha_1\alpha_2\alpha_3}\delta^{\beta_1\beta_2}-
		A_2(p_1\leftrightarrow p_2)\varepsilon^{p_2\alpha_1\alpha_2\alpha_3}\delta^{\beta_1\beta_2}
		\bigg]
	\end{aligned}
\end{equation}
\normalsize
with $A_1$ and $A_2$ given by eq$.$\eqref{eq:resultffconformi}.\\
The entire correlator is therefore determined {\em only by the anomalous coefficient} $a_2$ in \eqref{eew}.\\
We have also computed the correlator perturbatively at one-loop in free field theory and verified the agreement of the expression with the non-perturbative results obtained by imposing the conformal symmetry.  The explicit expressions of the form factors $A_1$ and $A_2$ have been given in \eqref{eq:pertresultff}. \\ The solutions of the conformal constraints, expressed in terms of 3K integrals $I_{5\{2,1,1\}}$ and $I_{4\{2,1,0\}}$, can be related to the ordinary one-loop master integrals $C_0$ and $B_0$ by \eqref{eqc}. They can be reconstructed using recursively the relations included in Appendix \ref{appmasterintegral}.

\section{Comments: non-renormalization of the $\langle AVV\rangle$ and $\langle TTJ_5 \rangle$ and the soft photon/graviton limits} \label{commnonrinormsec}
Before coming to our conclusions, we pause for few comments on the results of our paper, 
in relation to our previous study of the $\langle AVV\rangle$ chiral anomaly vertex, in a general CP-violating CFT \cite{Coriano:2023hts}.\\
In the case of the $\langle AVV\rangle$ vertex, the Adler-Bardeen theorem shows that the longitudinal part of the interaction is not affected by renormalization and therefore can be computed exactly just from the one-loop triangle diagram, being protected from perturbative corrections at higher orders.
This is not true for the transverse part of the same diagram, that satisfies an ordinary WI.
However, in \cite{Vainshtein:2002nv} it was pointed out that, in the kinematic limit where the momentum of one of the vector currents is vanishingly small, another non-renormalization theorem is valid.
Indeed, in that limit just two independent form factors are needed to fully describe the $\langle AVV\rangle$ correlator. One of these form factors is related to the axial anomaly and, therefore, it is not renormalized. The other form factor belongs to the transverse sector. \\
In \cite{Vainshtein:2002nv} it was shown that, due to helicity conservation in massless QCD, the two form factors are in fact proportional to each other, and so the non-renormalization of one of them implies that of the other. If the anomalous behaviour is identified with the exchange of an anomaly pole, 
that result relates the anomaly pole to the transverse part of the diagram, when one of the photons becomes soft. \\
Perturbative analysis of the diagram - in the most general kinematics - showed that at two-loops the entire diagram is indeed non renormalized  \cite{Jegerlehner:2005fs}, a feature that disappears at higher perturbative orders. Indeed, the authors of \cite{Mondejar:2012sz} found
non-vanishing corrections to the correlator at $O(\alpha^2_S)$.\\
The non-renormalization to all orders of a specific combination of the transverse form factors of the $\langle AVV\rangle$ was shown to hold in \cite{Knecht:2003xy}, in the chiral limit of QCD, since it equals the longitudinal form factor. The latter is, obviously, non-renormalized since the anomaly pole and its  residue are protected. Other combinations of purely transverse form factors were also shown to be non-renormalzed.   \\
In our anaysis \cite{Coriano:2023hts} we have shown - just using the conformal constraints - that such results indeed follow from conformal symmetry, once these are solved either in the most general kinematics 
or in the specific one required by Vainshtein's conjecture \cite{Vainshtein:2002nv}. 
Therefore, the breaking of the non-renormalizaton theorem for the entire vertex in QCD must originate from terms breaking conformal invariance and must be proportional to the QCD $\beta$ function.\\
In this paper we have verified that a similar connection between the longitudinal and the transverse part is present in the case of the $\langle TTJ_5\rangle$ correlator in the conformal limit, being both sectors proportional to the $a_2$ anomalous coefficient. \\
With these new indications, that follow quite closely the $\langle AVV\rangle $ case previously discussed by us, 
it would be interesting to test, at the perturbative level, if in the soft graviton limit a similar result holds for this correlator at all orders in perturbative QCD. We do expect that the higher order corrections will be proportional to the QCD $\beta$ function, therefore breaking the conformal symmetry. 

\section{Conclusions} 
We have presented an analysis of the gravitational anomaly vertex from the perspective of CFT in momentum space. We have shown how the vertex can be completely defined by the inclusion of a single anomaly pole together with the CWIs. This explicit analysis shows that reconstruction method formulated in the parity-even sector in the case of conformal anomaly correlators can be extended quite naturally to the parity-odd sector.  This provides a different and complementary perspective on the origin of anomalies and their related effective actions, which may account for such phenomena.
This extension highlights the intrinsic connection between these seemingly distinct sectors and suggests a unified framework for comprehending the origin of anomalies. It underscores the notion that anomalies, whether chiral, conformal, or supersymmetric, share a common underlying structure characterized by the presence of a single (anomaly) form factor, together with a specific tensor structure responsible for generating the anomaly.\\
 The approach does not rely on the explicit structure of the parity-odd current appearing in the correlator but, rather, on its symmetry properties. We have also shown that, similarly to previous dispersive analysis of the anomalous form factors for the $\langle TJJ\rangle $ and $\langle AVV\rangle$ diagrams,
the spectral density of the anomalous form factor of the $\langle TTJ_5\rangle$ satisfies a sum rule. The numerical value of the sum rule is fixed by the anomaly.  

\vspace{1.3cm}

\centerline{\bf Acknowledgements}
This work is partially supported by INFN within the Iniziativa Specifica QFT-HEP.  
The work of C. C. and S.L. is funded by the European Union, Next Generation EU, PNRR project "National Centre for HPC, Big Data and Quantum Computing", project code CN00000013. This work is partially supported by INFN, inziativa specifica {\em QG-sky} and by the the grant PRIN 2022BP52A MUR "The Holographic Universe for all Lambdas" Lecce-Naples. 
M.M.M. is supported by the European Research Council (ERC) under the European Union as Horizon 2020 research and innovation program (grant agreement No818066). 
C.C. thanks George Sterman and Peter Van Nieuwenhuizen for hospitality and discussions at the  C.N. Yang Institute for Theoretical Physics during a recent visit.  He thanks the Nuclear Theory Division at Brookhaven National Lab, and in particular Raju Venugopalan, Yoshitaka Hatta, Robert Pisarski, Andrej Tarasov and Xabier Feal  for hospitality and discussions. We finally thank Marco Guzzi of the Physics Dept. at KSU for discussions and George Zoupanos and Jan Kalinowski at the EISA, European Institute for Science and Their Applications at Corfu, Greece, for hospitality.

\appendix

\section{3K Integrals} \label{appendix:3kint}
The most general solution of the CWIs for our correlators can be written in terms of integrals involving a product of three Bessel functions, namely 3K integrals. In this appendix, we will illustrate such integrals and their properties. For a detailed review on the the topic, see \cite{Bzowski:2013sza,Bzowski:2015pba,Bzowski:2015yxv}.
\subsection{Definition and properties}
First, we recall the definition of the general 3K integral
\begin{equation}\label{eq:3kintdef}
	I_{\alpha\left\{\beta_1 \beta_2 \beta_3\right\}}\left(p_1, p_2, p_3\right)=\int d x x^\alpha \prod_{j=1}^3 p_j^{\beta_j} K_{\beta_j}\left(p_j x\right)
\end{equation}
where $K_\nu$ is a modified Bessel function of the second kind 
\begin{equation}
	K_\nu(x)=\frac{\pi}{2} \frac{I_{-\nu}(x)-I_\nu(x)}{\sin (\nu \pi)}, \qquad \nu \notin \mathbb{Z} \qquad\qquad I_\nu(x)=\left(\frac{x}{2}\right)^\nu \sum_{k=0}^{\infty} \frac{1}{\Gamma(k+1) \Gamma(\nu+1+k)}\left(\frac{x}{2}\right)^{2 k}
\end{equation}
with the property
\begin{equation}
	K_n(x)=\lim _{\epsilon \rightarrow 0} K_{n+\epsilon}(x), \quad n \in \mathbb{Z}
\end{equation}
The triple-K integral depends on four parameters: the power $\alpha$ of the integration variable $x$, and the three Bessel function indices $\beta_j$ . The arguments of the 3K integral are magnitudes of momenta $p_j$ with $j = 1, 2, 3$. One can notice the integral is invariant under the exchange $(p_j , \beta_j )\leftrightarrow (p_i , \beta_i )$.
We will also use the reduced version of the 3K integral defined as
\begin{equation}
	J_{N\left\{k_j\right\}}=I_{\frac{d}{2}-1+N\left\{\Delta_j-\frac{d}{2}+k_j\right\}}
\end{equation}
where we introduced the condensed notation $\{k_j \} = \{k_1, k_2, k_3 \}$.
The 3K integral satisfies an equation analogous to the dilatation equation with scaling degree
\begin{equation}
	\text{deg}\left(J_{N\left\{k_j\right\}}\right)=\Delta_t+k_t-2 d-N
\end{equation}
where 
\begin{equation}
	k_t=k_1+k_2+k_3,\qquad\qquad \Delta_t=\Delta_1+\Delta_2+\Delta_3
\end{equation}
From this analysis, it is simple to relate the form factors to the 3K integrals. Indeed, the dilatation WI of each from factor tells us that this needs to be written as a combination of integrals of the following type
\begin{equation}
	J_{N+k_t,\{k_1,k_2,k_3\}}
\end{equation}
where $N$ is the number of momenta that the form factor multiplies in the decomposition.
Let us now list some useful properties of 3K integrals
\begin{equation}\label{eq:3kprop}
	\begin{aligned}
		&\frac{\partial}{\partial p_n} J_{N\left\{k_j\right\}}  =-p_n J_{N+1\left\{k_j-\delta_{j n}\right\}} \\&
		J_{N\left\{k_j+\delta_{j n}\right\}}  =p_n^2 J_{N\left\{k_j-\delta_{j n}\right\}}+2\left(\Delta_n-\frac{d}{2}+k_n\right) J_{N-1\left\{k_j\right\}} \\&
		\frac{\partial^2}{\partial p_n^2} J_{N\left\{k_j\right\}}  =J_{N+2\left\{k_j\right\}}-2\left(\Delta_n-\frac{d}{2}+k_n-\frac{1}{2}\right) J_{N+1\left\{k_j-\delta_{j n}\right\}}, \\&
		K_n J_{N\left\{k_j\right\}}  \equiv\left(\frac{\partial^2}{\partial p_n^2}+\frac{\left(d+1-2 \Delta_n\right)}{p_n} \frac{\partial}{\partial p_n}\right) J_{N\left\{k_j\right\}}=J_{N+2\left\{k_j\right\}}-2 k_n J_{N+1\left\{k_j-\delta_{j n}\right\}},\\&
		K_{n m} J_{N\left\{k_j\right\}}\equiv (K_n-K_m)J_{N\left\{k_j\right\}} =-2 k_n J_{N+1\left\{k_j-\delta_{j n}\right\}}+2 k_m J_{N+1\left\{k_j-\delta_{j m}\right\}}
	\end{aligned}
\end{equation}

\subsection{Zero momentum limit} \label{appzerolimit3k}
When solving the secondary CWIs, it may be useful to perform a zero momentum limit.
In this subsection, we review the behaviour of the 3K integrals in the limit $p_3\rightarrow0$. In this limit, the momentum conservation gives
\begin{equation}
	p_{1}^{\mu}=-p_{2}^{\mu} \qquad \Longrightarrow \qquad p_1=p_2 \equiv p
\end{equation}
Assuming that $\alpha>\beta_t-1$ and $\beta_3>0$, we can write
\begin{equation}
	\lim _{p_3 \rightarrow 0} I_{\alpha\left\{\beta_j\right\}}\left(p, p, p_3\right)=p^{\beta_t-\alpha-1} \ell_{\alpha\left\{\beta_j\right\}}
\end{equation}
where 
\small
\begin{equation}
	\ell_{\alpha\left\{\beta_j\right\}}=\frac{2^{\alpha-3} \Gamma\left(\beta_3\right)}{\Gamma\left(\alpha-\beta_3+1\right)} \Gamma\left(\frac{\alpha+\beta_t+1}{2}-\beta_3\right) \Gamma\left(\frac{\alpha-\beta_t+1}{2}+\beta_1\right) \Gamma\left(\frac{\alpha-\beta_t+1}{2}+\beta_2\right) \Gamma\left(\frac{\alpha-\beta_t+1}{2}\right)
\end{equation}
\normalsize
We can derive similar formulas for the case $p_1\rightarrow0$ or $p_2\rightarrow0$ by 
considering the fact that 3K integrals are invariant under the exchange $(p_j , \beta_j )\leftrightarrow (p_i , \beta_i )$.

\subsection{Divergences and regularization}
The 3K integral defined in \eqref{eq:3kintdef} converges when 
\begin{equation}
	\alpha>\sum_{i=1}^3\left|\beta_i\right|-1 \quad ; \quad p_1, p_2, p_3>0
\end{equation}
If $\alpha$ does not satisfy this inequality, the integrals must be defined by an analytic continuation. 
The quantity
\begin{equation}
	\delta \equiv \sum_{j=1}^3\left|\beta_j\right|-1-\alpha
\end{equation}
is the expected degree of divergence.
However, when
\begin{equation}
	\alpha+1 \pm \beta_1 \pm \beta_2 \pm \beta_3=-2 k \quad, \quad k=0,1,2, \dots
\end{equation}
for some non-negative integer $k$ and any choice of the $\pm$ sign, the analytic continuation of the 3K integral generally has poles in the regularization parameter. 
Therefore, if the above condition is satisfied, we need to regularize the integrals. This can be done by shifting the parameters of the 3K integrals as
\begin{equation}
	I_{\alpha\left\{\beta_1, \beta_2, \beta_3\right\}} \rightarrow I_{\tilde{\alpha}\left\{\tilde{\beta}_1, \tilde{\beta}_2, \tilde{\beta}_3\right\}} \quad \Longrightarrow \quad J_{N\left\{k_1, k_2, k_3\right\}} \rightarrow J_{N+u \epsilon\left\{k_1+v_1 \epsilon, k_2+v_2 \epsilon, k_3+v_3 \epsilon\right\}}
\end{equation}
where
\begin{equation}
	\tilde{\alpha}=\alpha+u \epsilon \quad, \quad \tilde{\beta}_1=\beta_1+v_1 \epsilon \quad, \quad \tilde{\beta}_2=\beta_2+v_2 \epsilon \quad, \quad \tilde{\beta}_3=\beta_3+v_3 \epsilon
\end{equation}
or equivalently by considering
\begin{equation}
	d \rightarrow d+2 u \epsilon \quad ; \quad \Delta \rightarrow \Delta_i+\left(u+v_i\right) \epsilon
\end{equation}
In general, the regularisation parameters $u$ and $v_i$ are arbitrary. However, in certain cases, there may be some constraints on them. For simplicity, in this paper we consider the same $v_i=v$ for every $i$.

\subsection{3K integrals and Feynman integrals}
3K integrals are related to Feynman integrals in momentum space. The exact relations were first derived in \cite{Bzowski:2013sza,Bzowski:2015yxv}. Here we briefly show the results. Such expressions have been recently used in order to show the connection between the conformal analysis and the perturbative one for the $\langle AVV\rangle$ correlator \cite{Coriano:2023hts}.\\
Let $K_{d\{\delta_1\delta_2\delta_3\}}$ denote a massless scalar 1-loop 3-point momentum space integral
\begin{equation}
	K_{d\left\{\delta_1 \delta_2 \delta_3\right\}}=\int \frac{\mathrm{d}^d \boldsymbol{k}}{(2 \pi)^d} \frac{1}{k^{2 \delta_3}\left|\boldsymbol{p}_1-\boldsymbol{k}\right|^{2 \delta_2}\left|\boldsymbol{p}_2+\boldsymbol{k}\right|^{2 \delta_1}}
\end{equation}
Any such integral can be expressed in terms of 3K integrals and vice versa. For scalar integrals the relation reads
\begin{equation}
	K_{d\{\delta_1\delta_2\delta_3\}}=\frac{2^{4-\frac{3d}{2}}}{\pi^{\frac{d}{2}}}\times\frac{I_{\frac{d}{2}-1\{\frac{d}{2}+\delta_1-\delta_t,\frac{d}{2}+\delta_2-\delta_t,\frac{d}{2}+\delta_3-\delta_t \}}}{\Gamma(d-\delta_t)\Gamma(\delta_1)\Gamma(\delta_2)\Gamma(\delta_3)}
\end{equation}
where $\delta_t=\delta_1+\delta_2+\delta_3$. Its inverse reads
\begin{equation}
	\begin{aligned}
		& I_{\alpha\left\{\beta_1 \beta_2 \beta_3\right\}}=2^{3 \alpha-1} \pi^{\alpha+1} \Gamma\left(\frac{\alpha+1+\beta_t}{2}\right) \prod_{j=1}^3 \Gamma\left(\frac{\alpha+1+2 \beta_j-\beta_t}{2}\right) \\
		&\hspace{2.5cm} \times K_{2+2 \alpha,\left\{\frac{1}{2}\left(\alpha+1+2 \beta_1-\beta_t\right), \frac{1}{2}\left(\alpha+1+2 \beta_2-\beta_t\right), \frac{1}{2}\left(\alpha+1+2 \beta_3-\beta_t\right)\right\}}
	\end{aligned}
\end{equation}
where $\beta_t=\beta_1+\beta_2+\beta_3$. All tensorial massless 1-loop 3-point momentum-space integrals can also be expressed in terms of a number of 3K integrals when their tensorial structure is resolved by standard methods (for the exact expressions in this case see Appendix A.3 of \cite{Bzowski:2013sza}).

\section{Schouten identities} \label{appendix:Schouten}
In this section we derive the following minimal decomposition used when analyzing the special conformal constraint on the $\langle TTJ_5 \rangle$ correlator
\small
\begin{equation}\label{eq:decompscwi}
	\begin{aligned}
		0=\Pi^{\alpha_{1} \beta_{1}}_{\mu_{1} \nu_{1}}&\left({p}_{1}\right) \Pi^{\alpha_{2} \beta_{2}}_{\mu_{2} \nu_{2}}\left({p}_{2}\right) \pi^{\alpha_{3}}_{\mu_{3}}\left({p_3}\right) 
		\bigg(\mathcal{K}^\kappa 
		\left\langle T^{\mu_1 \nu_1}\left(p_1\right) T^{\mu_2 \nu_2}\left(p_2\right) J_5^{\mu_3}\left(p_3\right)\right\rangle
		\bigg)=
		\Pi^{\alpha_{1} \beta_{1}}_{\mu_{1} \nu_{1}}\left({p}_{1}\right) \Pi^{\alpha_{2} \beta_{2}}_{\mu_{2} \nu_{2}}\left({p}_{2}\right) \pi^{\alpha_{3}}_{\mu_{3}}\left({p_3}\right) \bigg[\\&
		p_1^\kappa \bigg(
		C_{11}\varepsilon^{p_1\mu_1\mu_2\mu_3}p_2^{\nu_1}p_3^{\nu_2}+
		C_{12}\varepsilon^{p_2\mu_1\mu_2\mu_3}p_2^{\nu_1}p_3^{\nu_2}+
		C_{13}\varepsilon^{p_1\mu_1\mu_2\mu_3}\delta^{\nu_1\nu_2}
		+C_{14}\varepsilon^{p_2\mu_1\mu_2\mu_3}\delta^{\nu_1\nu_2}\\&
		+C_{15}\varepsilon^{p_1p_2\mu_1\mu_2}p_2^{\nu_1}p_3^{\nu_2}p_1^{\mu_3}
		+C_{16}\varepsilon^{p_1p_2\mu_1\mu_2}\delta^{\nu_1\nu_2}p_1^{\mu_3}\bigg)\\&
		+p_2^\kappa \bigg(
		C_{21}\varepsilon^{p_1\mu_1\mu_2\mu_3}p_2^{\nu_1}p_3^{\nu_2}+
		C_{22}\varepsilon^{p_2\mu_1\mu_2\mu_3}p_2^{\nu_1}p_3^{\nu_2}+
		C_{23}\varepsilon^{p_1\mu_1\mu_2\mu_3}\delta^{\nu_1\nu_2}
		+C_{24}\varepsilon^{p_2\mu_1\mu_2\mu_3}\delta^{\nu_1\nu_2}\\&
		+C_{25}\varepsilon^{p_1p_2\mu_1\mu_2}p_2^{\nu_1}p_3^{\nu_2}p_1^{\mu_3}
		+C_{26}\varepsilon^{p_1p_2\mu_1\mu_2}\delta^{\nu_1\nu_2}p_1^{\mu_3}\bigg)\\&
		+\delta^{\kappa\mu_1}\bigg(
		C_{31}\varepsilon^{p_1\mu_2\mu_3\nu_1}p_3^{\nu_2}
		+C_{32}\varepsilon^{p_2\mu_2\mu_3\nu_1}p_3^{\nu_2}
		+C_{33}\varepsilon^{p_1p_2\mu_2\nu_1}p_1^{\mu_3}p_3^{\nu_2}
		+C_{34}\varepsilon^{p_1p_2\mu_2\mu_3}\delta^{\nu_1\nu_2}
		\bigg)
		\\&
		+\delta^{\kappa\mu_2}\bigg(
		C_{41}\varepsilon^{p_1\mu_1\mu_3\nu_2}p_2^{\nu_1}
		+C_{42}\varepsilon^{p_2\mu_1\mu_3\nu_2}p_2^{\nu_1}
		+C_{43}\varepsilon^{p_1p_2\mu_1\nu_2}p_1^{\mu_3}p_2^{\nu_1}
		+C_{44}\varepsilon^{p_1p_2\mu_1\mu_3}\delta^{\nu_1\nu_2}
		\bigg)
		\\&
		+C_{51}\varepsilon^{\kappa\mu_1\mu_2\mu_3}\delta^{\nu_1\nu_2}
		+C_{52}\varepsilon^{\kappa\mu_1\mu_2\mu_3}p_2^{\nu_1}p_3^{\nu_2}
		+C_{53}\varepsilon^{p_1\kappa\mu_1\mu_2}p_1^{\mu_3}\delta^{\nu_1\nu_2}
		+C_{54}\varepsilon^{p_2\kappa\mu_1\mu_2}p_1^{\mu_3}\delta^{\nu_1\nu_2}\bigg]
	\end{aligned}
\end{equation}
\normalsize
In order to determine such decomposition, first we have to write all the possible tensor structures that can appear in the equation. In particular the tensor related to the primary equations are 
\begin{equation}\label{eq:tenslistuno}
	\begin{gathered}
		\varepsilon^{p_1\mu_1\mu_2\mu_3}p_2^{\nu_1}p_3^{\nu_2}p_1^\kappa,\qquad
		\varepsilon^{p_1\mu_1\mu_2\mu_3}p_2^{\nu_1}p_3^{\nu_2}p_2^\kappa,\qquad
		\varepsilon^{p_2\mu_1\mu_2\mu_3}p_2^{\nu_1}p_3^{\nu_2}p_1^\kappa,\qquad
		\varepsilon^{p_2\mu_1\mu_2\mu_3}p_2^{\nu_1}p_3^{\nu_2}p_2^\kappa,\\
		\varepsilon^{p_1\mu_1\mu_2\mu_3}\delta^{\nu_1\nu_2}p_1^\kappa,\qquad
		\varepsilon^{p_1\mu_1\mu_2\mu_3}\delta^{\nu_1\nu_2}p_2^\kappa,\qquad
		\varepsilon^{p_2\mu_1\mu_2\mu_3}\delta^{\nu_1\nu_2}p_1^\kappa,\qquad
		\varepsilon^{p_2\mu_1\mu_2\mu_3}\delta^{\nu_1\nu_2}p_2^\kappa\\
		\varepsilon^{p_1p_2\mu_1\mu_2}p_2^{\nu_1}p_3^{\nu_2}p_1^{\mu_3}p_1^\kappa,\qquad
		\varepsilon^{p_1p_2\mu_1\mu_2}p_2^{\nu_1}p_3^{\nu_2}p_1^{\mu_3}p_2^\kappa,\qquad
		\varepsilon^{p_1p_2\mu_1\mu_2}\delta^{\nu_1\nu_2}p_1^{\mu_3}p_1^\kappa,\qquad
		\varepsilon^{p_1p_2\mu_1\mu_2}\delta^{\nu_1\nu_2}p_1^{\mu_3}p_2^\kappa,\\
		\varepsilon^{p_1p_2\mu_1\mu_2}p_2^{\nu_1}\delta^{\nu_2\mu_3}p_1^{\mu_3}p_1^\kappa,\qquad
		\varepsilon^{p_1p_2\mu_1\mu_2}p_2^{\nu_1}\delta^{\nu_2\mu_3}p_1^{\mu_3}p_2^\kappa,\qquad
		\varepsilon^{p_1p_2\mu_1\mu_2}p_1^{\nu_2}\delta^{\nu_1\mu_3}p_1^{\mu_3}p_2^\kappa,\qquad
	\end{gathered}
\end{equation}
and similar ones for the secondary. 
However, not all of these tensors are independent. Some of these tensors can be rewritten in terms of each others. We then need to find a set of tensors that form a minimal decomposition. We will illustrate a couple of cases of Schouten identities needed for this purpose. For example we consider the equation
\begin{equation} \label{eq:sciduno}
	0=\varepsilon^{[p_1p_2\mu_1\mu_2}\delta^{\mu_3]}_\alpha
\end{equation}
which can be contracted with $p_{1}^\alpha$ and $p_{2}^\alpha$, obtaining
\begin{equation}
	\begin{aligned}
		\Pi^{\alpha_{1} \beta_{1}}_{\mu_{1} \nu_{1}}\left({p}_{1}\right) \Pi^{\alpha_{2} \beta_{2}}_{\mu_{2} \nu_{2}}\left({p}_{2}\right) \pi^{\alpha_{3}}_{\mu_{3}}\left({p_3}\right) &\bigg[ \varepsilon^{p_1 p_2\mu_1\mu_3} p_3^{\mu_2}\bigg]=\Pi^{\alpha_{1} \beta_{1}}_{\mu_{1} \nu_{1}}\left({p}_{1}\right) \Pi^{\alpha_{2} \beta_{2}}_{\mu_{2} \nu_{2}}\left({p}_{2}\right) \pi^{\alpha_{3}}_{\mu_{3}}\left({p_3}\right) \bigg[\\&
		\frac{1}{2} \varepsilon^{p_1\mu_1\mu_2\mu_3} \left(-p_1^2-p_2^2+p_3^2\right)-\varepsilon^{p_1 p_2\mu_1\mu_2} p_1^{\mu_3}-\varepsilon^{p_2\mu_1\mu_2
			\mu_3} p_1^2\bigg],\\
		\Pi^{\alpha_{1} \beta_{1}}_{\mu_{1} \nu_{1}}\left({p}_{1}\right) \Pi^{\alpha_{2} \beta_{2}}_{\mu_{2} \nu_{2}}\left({p}_{2}\right) \pi^{\alpha_{3}}_{\mu_{3}}\left({p_3}\right)& \bigg[
		\varepsilon^{p_1 p_2\mu_2\mu_3} p_2^{\mu_1}\bigg]=\Pi^{\alpha_{1} \beta_{1}}_{\mu_{1} \nu_{1}}\left({p}_{1}\right) \Pi^{\alpha_{2} \beta_{2}}_{\mu_{2} \nu_{2}}\left({p}_{2}\right) \pi^{\alpha_{3}}_{\mu_{3}}\left({p_3}\right) \bigg[\\&
		\varepsilon^{p_1\mu_1\mu_2\mu_3} p_2^2+\varepsilon^{p_1 p_2\mu_1\mu_2} p_1^{\mu_3}-\frac{1}{2} \varepsilon^{p_2\mu_1\mu_2\mu_3} \left(-p_1^2-p_2^2+p_3^2\right)\bigg].
	\end{aligned}	
\end{equation}
We then consider the identity
\begin{equation}
	0=\varepsilon^{[p_1p_2\mu_1\mu_2}\delta^{\kappa]}_\alpha
\end{equation}
which can be contracted with $p_{1}^\alpha$ and $p_{2}^\alpha$, obtaining
\begin{equation}
	\begin{aligned}
		\Pi^{\alpha_{1} \beta_{1}}_{\mu_{1} \nu_{1}}\left({p}_{1}\right)& \Pi^{\alpha_{2} \beta_{2}}_{\mu_{2} \nu_{2}}\left({p}_{2}\right) \pi^{\alpha_{3}}_{\mu_{3}}\left({p_3}\right) \bigg[\varepsilon^{p_1 p_2\kappa\mu_1} p_3^{\mu_2}\bigg]=\\&
		\Pi^{\alpha_{1} \beta_{1}}_{\mu_{1} \nu_{1}}\left({p}_{1}\right) \Pi^{\alpha_{2} \beta_{2}}_{\mu_{2} \nu_{2}}\left({p}_{2}\right) \pi^{\alpha_{3}}_{\mu_{3}}\left({p_3}\right) \bigg[ -\frac{1}{2} \varepsilon^{p_1\kappa\mu_1\mu_2} \left(-p_1^2-p_2^2+p_3^2\right)+\varepsilon^{p_1 p_2\mu_1\mu_2} p_1^\kappa+\varepsilon^{p_2\kappa\mu_1\mu_2} p_1^2\bigg],\\
		\Pi^{\alpha_{1} \beta_{1}}_{\mu_{1} \nu_{1}}\left({p}_{1}\right)& \Pi^{\alpha_{2} \beta_{2}}_{\mu_{2} \nu_{2}}\left({p}_{2}\right) \pi^{\alpha_{3}}_{\mu_{3}}\left({p_3}\right) \bigg[
		\varepsilon^{p_1 p_2\kappa\mu_2} p_2^{\mu_1}\bigg]=\\&
		\Pi^{\alpha_{1} \beta_{1}}_{\mu_{1} \nu_{1}}\left({p}_{1}\right) \Pi^{\alpha_{2} \beta_{2}}_{\mu_{2} \nu_{2}}\left({p}_{2}\right) \pi^{\alpha_{3}}_{\mu_{3}}\left({p_3}\right) \bigg[ -\varepsilon^{p_1\kappa\mu_1\mu_2} p_2^2+\varepsilon^{p_1 p_2\mu_1\mu_2} p_2^\kappa+\frac{1}{2} \varepsilon^{p_2\kappa\mu_1\mu_2} \left(-p_1^2-p_2^2+p_3^2\right)\bigg]
	\end{aligned}
\end{equation}
The analysis of the remaining contraints is rather involved, but follows the steps outlined above. 
In the end, after considering all the possible Schouten identities, one finds a minimal set of independent structures as expressed in eq$.$ \eqref{eq:decompscwi}.
\section{Vertices} \label{appvertices}
In this section we list the explicit expression of all the vertices needed for the perturbative analysis of the $\langle TTJ_5\rangle$ correlator. The momenta of the gravitons and the axial boson are all incoming as well as the momentum indicated with $k_1$. The momentum $k_2$ instead is outgoing. In order to simplify the notation, we introduce the tensor components
\begin{equation}
	\begin{aligned}
		&A^{\mu\nu\rho\sigma}\equiv g^{\mu  \nu } g^{\rho  \sigma }-\frac{1}{2} \left(g^{\mu  \rho } g^{\nu  \sigma }+g^{\mu  \sigma } g^{\nu  \rho }\right)
		\\&
		B^{\mu\nu\rho\sigma\alpha\beta}\equiv g^{\alpha  \beta } g^{\mu  \nu } g^{\rho  \sigma }-g^{\alpha  \beta } \left(g^{\mu  \rho } g^{\nu  \sigma }+g^{\mu  \sigma } g^{\nu  \rho }\right)
		\\&
		C^{\mu\nu\rho\sigma\alpha\beta} \equiv \frac{1}{2} g^{\mu  \nu } \left(g^{\alpha  \rho } g^{\beta  \sigma }+g^{\alpha  \sigma } g^{\beta  \rho }\right)+\frac{1}{2} g^{\rho  \sigma } \left(g^{\alpha  \mu } g^{\beta  \nu }+g^{\alpha  \nu } g^{\beta  \mu }\right)
		\\&
		D^{\mu\nu\rho\sigma\alpha\beta}\equiv \frac{1}{2} \left(g^{\alpha  \sigma } g^{\beta  \mu } g^{\nu  \rho }+g^{\alpha  \rho } g^{\beta  \mu } g^{\nu  \sigma }+g^{\alpha  \sigma } g^{\beta  \nu } g^{\mu  \rho }+g^{\alpha  \rho } g^{\beta  \nu } g^{\mu  \sigma }\right)+\\&
		\qquad\qquad \qquad 
		\frac{1}{4} \left(g^{\alpha  \mu } g^{\beta  \sigma } g^{\nu  \rho }+g^{\alpha  \mu } g^{\beta  \rho } g^{\nu  \sigma }+g^{\alpha  \nu } g^{\beta  \sigma } g^{\mu  \rho }+g^{\alpha  \nu } g^{\beta  \rho } g^{\mu  \sigma }\right)
		\\&
		G^{\alpha \beta \gamma}\equiv \gamma^{\alpha }\gamma^{\beta }\gamma^{\gamma }-\gamma^{\beta 	}\gamma^{\alpha }\gamma^{\gamma }+\gamma^{\gamma }\gamma^{\alpha }\gamma^{\beta }-\gamma ^{\gamma }\gamma^{\beta }\gamma^{\alpha }
	\end{aligned}
\end{equation}
The vertices can then be written as

\begin{minipage}{0.28\textwidth}
	\includegraphics[width=\linewidth]{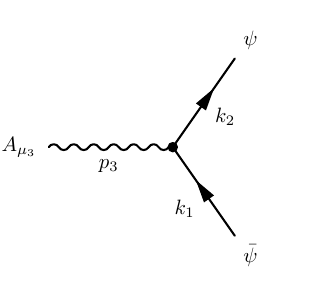}
\end{minipage}%
\begin{minipage}{0.5\textwidth}
	\begin{equation}
		\begin{aligned}
			V^{\mu_3}_{A\bar{\psi}\psi}=-i g \gamma^{\mu_3} \gamma_5  \nn
		\end{aligned}
	\end{equation}
\end{minipage}

\begin{minipage}{0.3\textwidth}
	\includegraphics[width=\linewidth]{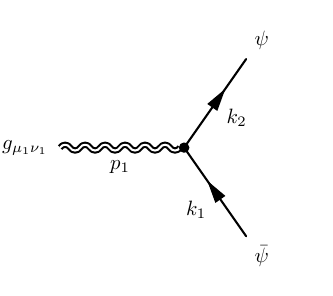}
\end{minipage}%
\begin{minipage}{0.6\textwidth}
	\begin{equation}
		\begin{aligned}
			V^{\mu_1 \nu_1}_{g \bar{\psi}\psi}=\frac{i}{4}A^{\mu_1 \nu_1 \rho \sigma}(k_1+k_2)_\rho \gamma_\sigma  \nn
		\end{aligned}
	\end{equation}
	\normalsize
\end{minipage}

\begin{minipage}{0.25\textwidth}
	\includegraphics[scale=0.9]{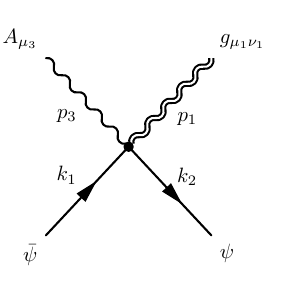}
\end{minipage}%
\begin{minipage}{0.68\textwidth}
	\begin{equation}
		\begin{aligned}
			V^{\mu_1 \nu_1 \mu_3}_{gA\bar{\psi}\psi}=-\frac{i g}{2}A^{\mu_1 \nu_1 \mu_3 \rho} \gamma_\rho \gamma_5 \nn
		\end{aligned}
	\end{equation}
	\normalsize
\end{minipage}

\begin{minipage}{0.35\textwidth}
	\includegraphics[scale=0.87]{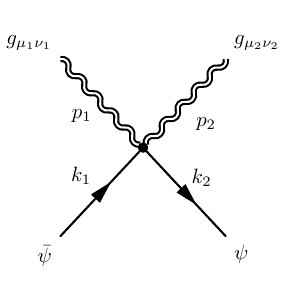}
\end{minipage}%
\begin{minipage}{0.55\textwidth}
	\begin{equation}
		\begin{aligned}
			&V^{\mu_1\nu_1\mu_2\nu_2}_{gg\bar{\psi}\psi}=\\&\quad -\frac{i }{8}\left(B^{\mu_1\nu_1\mu_2\nu_2 \alpha \beta}-C^{\mu_1\nu_1\mu_2\nu_2 \alpha \beta}+D^{\mu_1\nu_1\mu_2\nu_2 \alpha \beta}\right)\gamma_\alpha(k_1+k_2)_\beta\\&\quad 
			+\frac{i}{128}G^{\alpha \beta \gamma } A^{\mu_1\nu_1  \gamma \rho}p_2^\sigma \,\, \times \\&\qquad
			\left(g^{\alpha  \mu_2} g^{\beta  \sigma } g^{\nu_2 \rho }+g^{\alpha  \nu_2} g^{\beta  \sigma } g^{\mu_2 \rho }-g^{\alpha  \sigma } g^{\beta  \nu_2} g^{\mu_2 \rho }-g^{\alpha  \sigma } g^{\beta  {\mu_2}} g^{\nu_2 \rho }\right) \nn
		\end{aligned}
	\end{equation}
\end{minipage}

\begin{minipage}{0.35\textwidth}
	\includegraphics[scale=0.87]{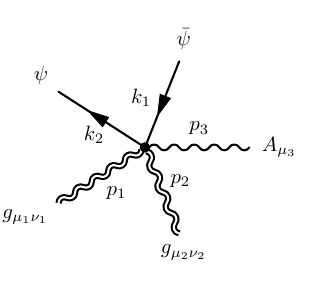}
\end{minipage}%
\begin{minipage}{0.6\textwidth}
	\begin{equation}
		\begin{aligned}
			&V^{\mu_1\nu_1\mu_2\nu_2\mu_3}_{ggA\bar{\psi}\psi}=
			\\&\qquad 
			-\frac{ig}{4} \left(B^{\mu_1\nu_1\mu_2\nu_2\mu_3 \lambda}-C^{\mu_1\nu_1\mu_2\nu_2\mu_3 \lambda}+D^{\mu_1\nu_1\mu_2\nu_2\mu_3 \lambda}\right) \gamma_{\lambda}\gamma_5\nn
		\end{aligned}
	\end{equation}
	\normalsize
\end{minipage}

\section{Master integrals} 
\label{appmasterintegral}
In this section we summarize some important relations regarding the master integrals $B_0$ and $C_0$. They are defined by the following expressions in the Euclidean space\footnote{In the Minkoski space, the prefactor of the integrals is $1/(i \pi^2)$.}
\begin{equation}
	\begin{aligned}
		&B_0(p_i^2)  \equiv \frac{1}{  \pi^2}  \int {d^d l} \frac{1}{l^2\left(l-p_i\right)^2} = \frac{1}{\varepsilon }+\log\left( - \frac{\mu^2 }{\pi  {p}^2_i}\right)-\gamma +2 
	\end{aligned}
\end{equation}
and
\begin{equation}
	\begin{aligned}
		&C_{0}(p_1^2,p_2^2,p_3^2) \equiv \frac{1}{  \pi^2}  \int d^d l \frac{1}{l^2(l-p_1)^2(l+p_2)^2}=\\
		&\quad 
		\frac{1}{\sqrt{\lambda}}\, \Bigg[\text{Li}_2\left(-\frac{-p_1^2+p_2^2+p_3^2+\sqrt{\lambda }}{p_1^2-p_2^2-p_3^2+\sqrt{\lambda }}\right)+\text{Li}_2\left(-\frac{p_1^2-p_2^2+p_3^2+\sqrt{\lambda }}{-p_1^2+p_2^2-p_3^2+\sqrt{\lambda }}\right)
		-\text{Li}_2\left(\frac{p_1^2+p_2^2-p_3^2-\sqrt{\lambda }}{p_1^2+p_2^2-p_3^2+\sqrt{\lambda }}\right)\\&\qquad
		+\text{Li}_2\left(-\frac{p_1^2+p_2^2-p_3^2+\sqrt{\lambda }}{-p_1^2-p_2^2+p_3^2+\sqrt{\lambda }}\right)-\text{Li}_2\left(\frac{p_1^2-p_2^2+p_3^2-\sqrt{\lambda }}{p_1^2-p_2^2+p_3^2+\sqrt{\lambda }}\right)-\text{Li}_2\left(\frac{-p_1^2+p_2^2+p_3^2-\sqrt{\lambda }}{-p_1^2+p_2^2+p_3^2+\sqrt{\lambda }}\right)\Bigg]
	\end{aligned}
\end{equation}
with $\lambda\equiv \lambda(p_1,p_2,p_3)$ defined in \eqref{Kallen}.
By acting with derivatives on such integrals one finds \cite{Coriano:2018bbe} 
\begin{equation}
	\frac{\partial}{\partial p_i} B_0\left(p_i^2\right)=\frac{(d-4)}{p_i} B_0\left(p_i^2\right)
\end{equation}
and
\begin{equation}
	\begin{aligned}
		&\frac{\partial}{\partial p_1} C_0=  \frac{1}{\lambda p_1}\Bigg\{2(d-3)\bigg[\left(p_1^2+p_2^2-p_3^2\right) B_0\left(p_2^2\right)+\left(p_1^2-p_2^2+p_3^2\right) B_0\left(p_3^2\right)-2 p_1^2 B_0(p_1^2)\bigg] \\
		&\qquad\quad+  {\bigg[(d-4)\left(p_2^2-p_3^2\right)^2-(d-2) p_1^4+2 p_1^2\left(p_2^2+p_3^2\right)\bigg] C_0\left(p_1^2, p_2^2, p_3^2\right)\Bigg\} } \\&
		\frac{\partial}{\partial p_2} C_0=  \frac{1}{\lambda p_2}\Bigg\{2(d-3)\bigg[\left(p_1^2+p_2^2-p_3^2\right) B_0(p_1^2)+\left(p_2^2+p_3^2-p_1^2\right) B_0\left(p_3^2\right)-2 p_2^2 B_0\left(p_2^2\right)\bigg] \\
		&\qquad\quad+  {\bigg[2 p_3^2\left(p_2^2-(d-4) p_1^2\right)+\left(p_1^2-p_2^2\right)\left((d-4) p_1^2+(d-2) p_2^2\right)+(d-4) p_3^4\bigg] C_0\left(p_1^2, p_2^2, p_3^2\right)\Bigg\} } \\&
		\frac{\partial}{\partial p_3} C_0=  \frac{1}{\lambda p_3}\Bigg\{2(d-3)\bigg[\left(p_1^2-p_2^2+p_3^2\right) B_0(p_1^2)+\left(p_2^2+p_3^2-p_1^2\right) B_0\left(p_2^2\right)-2 p_3^2 B_0\left(p_3^2\right)\bigg] \\
		&\qquad\quad+\bigg[(d-4)\left(p_1^2-p_2^2\right)^2-(d-2) p_3^4+2 p_3^2\left(p_1^2+p_2^2\right)\bigg] C_0\left(p_1^2, p_2^2, p_3^2\right)\Bigg\} 
	\end{aligned}
\end{equation}

\providecommand{\href}[2]{#2}\begingroup\raggedright\endgroup

\end{document}